\documentclass[11pt,english]{article}
\usepackage{geometry}
\usepackage{longtable}
\usepackage{booktabs}
\usepackage{multirow}
\usepackage{setspace}
\usepackage[authoryear]{natbib}
\usepackage[english]{babel}
\usepackage{amsmath, amsthm, amssymb, amscd, amsfonts, bm, bbm, mathrsfs}   
\usepackage{graphicx}
\usepackage{enumerate}
\usepackage[shortlabels]{enumitem}
\usepackage{url} 
\usepackage[colorlinks]{hyperref}
\usepackage{xr-hyper}
\usepackage[capitalize,nameinlink,noabbrev]{cleveref} 
\usepackage[algo2e,linesnumbered,lined,ruled]{algorithm2e}

\usepackage[bottom]{footmisc}
\usepackage[dvipsnames]{xcolor}
\hypersetup{
	breaklinks,
	linkcolor=blue,
	urlcolor=blue,
	anchorcolor=blue,
	citecolor=blue
}

\makeatletter

\theoremstyle{plain}

\@ifundefined{date}{}{\date{}}
\usepackage{babel}
\bibpunct{(}{)}{,}{autheryear}{,}{,}

\allowdisplaybreaks

\usepackage{tikz}
\usetikzlibrary{fit,positioning}

\allowdisplaybreaks

\newcommand*{\indep}{%
  \mathbin{%
    \mathpalette{\@indep}{}%
  }%
}
\newcommand*{\nindep}{%
  \mathbin{
    \mathpalette{\@indep}{\not}
  }%
}
\newcommand*{\@indep}[2]{%
  \sbox0{$#1\perp\m@th$}
  \sbox2{$#1=$}
  \sbox4{$#1\vcenter{}$}
  \rlap{\copy0}
  \dimen@=\dimexpr\ht2-\ht4-.2pt\relax
  \kern\dimen@
  {#2}%
  \kern\dimen@
  \copy0 
}

\newtheorem{thm}{Theorem}

\newtheorem{defn}{Definition}

\newtheorem{rem}{Remark}

\newtheorem{eg}{Example}

\newcommand{\dsp}{\displaystyle} 

\newcommand{\argmin}{\operatornamewithlimits{argmin}} 

\makeatletter
\newcommand*{\rom}[1]{\expandafter\@slowromancap\romannumeral #1@}
\newcommand{\HUGE}{\@setfontsize\Huge{40}{50}}   
\makeatother

\makeatletter
\newcommand{\labelcustom}[2]{%
	\protected@write \@auxout {}{\string \newlabel {#1}{{#2}{\thepage}{#2}{#1}{}} }%
	\hypertarget{#1}{#2}
}
\makeatother

\makeatletter
\newcommand{\labeltext}[3][]{%
	\@bsphack%
	\csname phantomsection\endcsname
	\def\tst{#1}%
	\def\refmarkup{}%
	\ifx\tst\empty\def\@currentlabel{\refmarkup{#2}}{\label{#3}}%
	\else\def\@currentlabel{\refmarkup{#1}}{\label{#3}}\fi%
	\@esphack%
	\labelmarkup{#2}
}
\makeatother

\makeatletter
\newcommand{\bianca}{\renewcommand\NAT@open{[}\renewcommand\NAT@close{]}}
\makeatother

\newcommand{\iid}{\overset{\mathsf{iid}}{\sim}} 

\newcommand{\pr}{\mathsf{P}}
\newcommand{\eo}{\mathsf{E}}
\newcommand{\cov}{\mathsf{cov}}

\newcommand{\nd}{\mathsf{N}}

\newcommand{\ap}{\alpha} 
\newcommand{\g}{\gamma} 
\newcommand{\ga}{\Gamma} 
\newcommand{\e}{\varepsilon} 
\newcommand{\Oa}{\Omega} 
\newcommand{\ld}{\lambda} 
\newcommand{\Ld}{\Lambda} 

\newcommand{\B}{\mathbb{B}} 
\newcommand{\HH}{\mathbb{H}} 
\newcommand{\I}{\mathbb{I}} 
\newcommand{\R}{\mathbb{R}} 
\newcommand{\SSS}{\mathbb{S}} 
\newcommand{\V}{\mathbb{V}} 

\newcommand{\aaa}{\mathcal{A}}	
\newcommand{\bb}{\mathcal{B}}	
\newcommand{\ff}{\mathcal{F}}	
\newcommand{\fgg}{\mathcal{G}}	
\newcommand{\hh}{\mathcal{H}}	
\newcommand{\ii}{\mathcal{I}}	
\newcommand{\nn}{\mathcal{N}}	
\newcommand{\oo}{\mathcal{O}}	
\newcommand{\rr}{\mathcal{R}}	
\newcommand{\uu}{\mathcal{U}}	
\newcommand{\vv}{\mathcal{V}}	
\newcommand{\xx}{\mathcal{X}}	
\newcommand{\yy}{\mathcal{Y}}	

\let\oldnl\nl
\newcommand{\nlnonumber}{\renewcommand{\nl}{\let\nl\oldnl}}
\newcommand{\KwTune}{\KwSty{Parameters:}}

\renewcommand{\eqref}[1]{(\ref{#1})}

\begin{document}

	\renewcommand{\sectionautorefname}{Section}
	\renewcommand{\subsectionautorefname}{Section}
	\renewcommand{\subsubsectionautorefname}{Section}
	\renewcommand{\algorithmautorefname}{Algorithm}
	
\title{
	Regularized Halfspace Depth for Functional Data
}
\author{
	Hyemin Yeon\thanks{
		Department of Statistics, North Carolina State University, North Carolina 27695, U.S.A. Email: hyeon@ncsu.edu.
		}, 
	Xiongtao Dai\thanks{
		Division of Biostatistics, University of California, Berkeley
		}, 
	and Sara Lopez-Pintado\thanks{
		Department of Health Sciences, Northeastern University
	} 
}

\maketitle
\begin{abstract}
	
	Data depth is a powerful nonparametric tool originally proposed to rank multivariate data from center outward.  
	In this context, one of the most archetypical depth notions is Tukey's halfspace depth. 
	In the last few decades notions of depth have also been proposed for functional data. 
	However, Tukey's depth cannot be extended to handle functional data because of its degeneracy. 
	Here, we propose a new halfspace depth for functional data which avoids degeneracy by regularization. 
	The halfspace projection directions are constrained to have a small reproducing kernel Hilbert space norm. 
	Desirable theoretical properties of the proposed depth, such as isometry invariance, maximality at center, monotonicity relative to a deepest point, upper semi-continuity, and consistency are established. 
	Moreover, the regularized halfspace depth can rank functional data with varying emphasis in shape or magnitude, depending on the regularization. 
	A new outlier detection approach is also proposed, which is capable of detecting both shape and magnitude outliers. 
	It is applicable to trajectories in $L^2$, a very general space of functions that include non-smooth trajectories. 
	Based on extensive numerical studies, our methods are shown to perform well in terms of detecting outliers of different types. 
	Three real data examples showcase the proposed depth notion. 


\end{abstract}



\section{Introduction}


\subsection{Backgrounds and degeneracy issue}

In the last several decades, data depth methods for multivariate data have been developed and are shown to be a powerful tool for both exploratory data analysis and robust statistics. 
Many different depth notions have been proposed, such as Tukey's halfspace depth \citep{tukey75}, simplicial depth \citep{Liu90}, and Mahalanobis depth \citep{Liu92}.
These multivariate depth functions have not only been well-investigated theoretically \citep{DG92, ZS00} but have also been applied for handling different statistical problems \citep{
	YS97, 
	RRT99, LCL12}.

\begin{figure}[b!]
	\centering
	\includegraphics[width=0.69\linewidth]{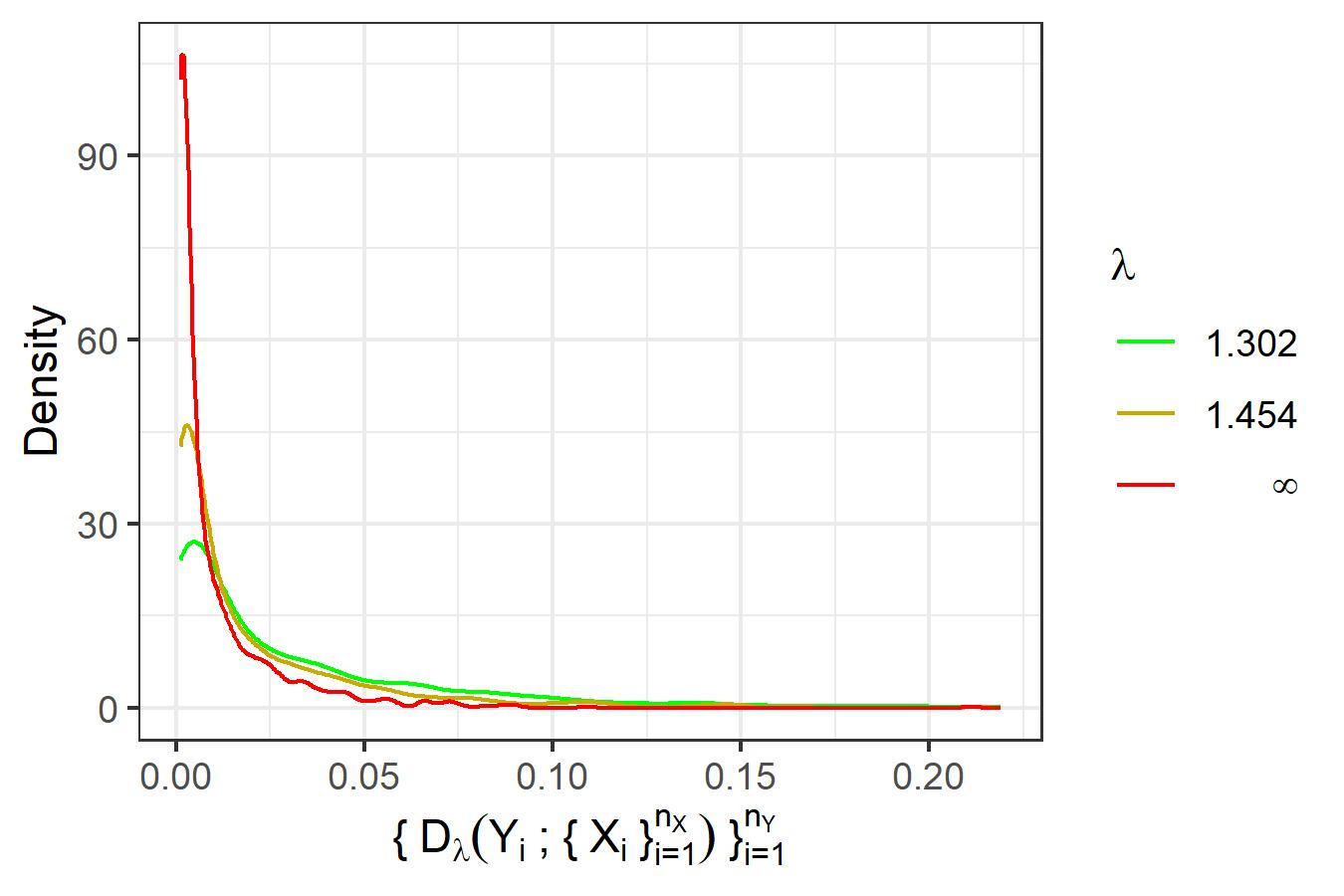}
	\caption{
		Distributions of the original Tukey's halfspace depth values and the proposed regularized halfspace depth with different regularization parameters $\ld \in \{ 1.302, 1.454 \}$.
		The depth values are w.r.t. $\xx_n \equiv \{X_i\}_{i=1}^{n_X}$ and evaluated at the points in $\yy_n \equiv \{Y_i\}_{i=1}^{n_Y}$, where the observations are i.i.d.\ realizations from a Gaussian process and $n_X = n_Y = 1000$.
		Tukey's depth corresponds to $\ld = \infty$.
	}
	\label{figSim1}
\end{figure}

In particular, Tukey's halfspace depth has been popular due to its many desirable properties \citep{ZS00} and the robustness of its depth median \citep{DG92}.
However, its generalization to functional data inevitably encounters a degeneracy issue, namely, a naive extension of Tukey's depth has zero depth values almost surely. 
Intuitively, this is due to the infinite-dimensionality of the functional space, where the set of projections is ``too rich'' and thus Tukey's depth defined from the most extreme projection is too small.
A numerical illustration of the degeneracy phenomenon in practice is shown in \autoref{figSim1}, where the naive direct extension of Tukey's depth to functional data (red curve) assigns zero depth to the majority of observations, preventing depth comparisons. 
The degeneracy was first discovered by \cite{DGC11} and further investigated by \cite{KZ13, CC14a}. 
See Table~1 of \cite{GN17} for a summary of the degeneracy issues in this and other existing functional depth notions.
Tukey's halfspace depth has therefore been practically dismissed from consideration for functional data due to the failure of the naive generalization.


\subsection{Regularized halfspace depth}

In this work, we propose a new notion of functional depth called the regularized halfspace depth (RHD). 
We resolve the degeneracy issue 
by restricting the set of projection directions in the definition of the halfspace depth 
to the set of directions with reproducing kernel Hilbert space (RKHS) norm less than a given positive number $\ld \in (0, \infty)$.
Indeed, the proposed RHD obtained in this fashion is non-degenerate over a general class of random functions.
This allows for depth-based rankings if, for example, the underlying distribution of the random element is Gaussian.
\autoref{figSim1} illustrates that the RHD is free of the degeneracy issue as shown by the green and yellow curves for small and large $\lambda$ values; details about this simulation setup can be found in Section~S2.1 in the Supplemental Materials.
Smaller $\lambda$ results in greater regularization, in which case the distribution of the depth values become more positive and dispersed, while the original Tukey's halfspace depth, corresponding to the red curve with $\ld=\infty$, exhibits a degenerate behavior, since almost all observations have zero depth.


A related but highly different projection-based Tukey's depth for functional data has been considered by \cite{CN08}, who proposed the random Tukey depth which defines the depth as the most severe extremity as seen along a \emph{fixed number} of random projections. 
The number of projections should not diverge to infinity because otherwise the random Tukey depth will approach Tukey's halfspace depth, which degenerates given functional observations.  
In contrast, our approach targets to utilize \emph{infinitely many} projections and thus is more sensitive against different types of extremeness; the degeneracy issue is resolved through regularizing the projections.
These approaches both demonstrate the necessity of applying restrictions when defining functional depth through the infimum over projection directions.

In addition to non-degeneracy, 
we establish other desirable depth properties for the RHD 
in analogy to the ones for multivariate and functional data postulated by \cite{ZS00}, \cite{NB16}, and  \cite{GN17}, respectively. 
We also include topological properties of the depth level sets.
A uniform consistency of the sample RHD for the population RHD is verified over a totally bounded subset.
A version of the Glivenko--Cantelli theorem is derived for a class of infinite-dimensional halfspaces as a by-product.
In all, our new depth satisfies all of the desirable properties considered by \cite{GN17}, unlike most other existing functional depths.

A prominent feature of the RHD is its flexibility in ranking curves based on different shape features controlled by the regularization parameter $\ld$: RHD with a larger $\ld$ emphasizes higher-frequency shape differences in the samples, while a smaller $\ld$ emphasizes overall magnitude. 
The definition of ``shape'' is data-driven and comes from the modes of variation in the observed data.
This feature leads to our new definitions of extremeness for curves informed by data, which is utilized and demonstrated in other applications such as outlier detection method and will be introduced next.


\subsection{Functional outlier detection}

In recent years, substantial development of outlier detection methods in functional data have been accomplished by using notions of functional depth.
The functional boxplot proposed by \cite{SG11} is an extension of the original boxplot \cite[cf.][]{tukey77}, where functional central region and whiskers are constructed based on a functional depth method, in particular the modified band depth (MBD) \citep{LR09}. 
Outlier detection methods for functions taking values in a multivariate space have also been considered, by e.g., \cite{HRS15}.


Trajectories outlying in only shape but not at any given time point are challenging to detect; \cite{NGH17} focused on the detection of this type of outliers by applying data depths to the joint distribution of a set of two or three time points altogether.
\cite{DMSG20} proposed to apply transformations sequentially to the functional data to reveal shape outliers that are not easily detected by other methods. 
Alternative outlier detection methods based on functional data depth include functional bagplots \citep{HS10}, the outliergram \citep{AR14}, and directional outlyingness \citep{DG18}. 

Unlike magnitude outliers that tend to have an extreme average value over the support, shape outliers are hard to define in the first place because there is vast possibility of outlyingness in shape, in terms of, e.g., jumps, peaks, smoothness, trends etc. 
A shape outlier can often be inlying pointwise but has a different overall pattern than the majority of the sample curves over the support.
A major benefit of the proposed RHD is that it captures different modes of variation in the sample $\xx_n \equiv \{X_i\}_{i=1}^n$ depending on the choice of the regularization parameter $\ld$.
We propose a \emph{data-adaptive} definition of shape outlyingness and establish a novel shape outlier detection method based on the proposed RHD.
The method uses a large number of randomly drawn directions to approximate the sample RHD and applies the univariate boxplot to the data functions projected onto those directions.
Acknowledging the reviewers’ recommendations, we have also considered phase variation and phase outliers.
Namely, the observed curves exhibit variability in the x-axis, for example, when each curve is shifted (or delayed) compared to the others.
As provided in the numerical studies, the proposed outlier detection method well detects different types of extremeness/outlyingness in shape, phase, as well as in magnitude.

Our methods are shown to be practical and useful with different real data examples: medfly, world population growth, and sea surface temperature datasets.
Due to the complexity of the first dataset, it has barely been studied \cite[cf.][]{SL21} in the literature of functional outlier detection, while the other two datasets have been analyzed a few times previously \cite[cf.][]{SG11, NGH17, DMSG20}.
Our data illustrations demonstrate that the proposed method is relevant for analyzing both complicated and rough trajectories as well as relatively simple smooth curves.

\subsection{Contributions and outline of the paper}

Next we highlight our main contributions while summarizing the organization of the paper.
\autoref{sec2} defines and studies the proposed RHD, 
which is a novel, flexible and useful depth notion for functional data; 
it is shown in \autoref{ssec_2_1} that the proposed RHD is non-degenerate over a large class of random functions.
We then establish desirable depth properties of the RHD in Sections~\ref{ssec_2_2}-\ref{ssec_2_3} 
and provide a theoretically reasonable and practical algorithm to compute it in \autoref{ssec_2_4};
some intermediate results such as a Glivenko-Cantelli theorem are new in the functional depth literature.
The choice of relevant tuning parameters are discussed in \autoref{ssec_2_5}. 
\autoref{sec3} is devoted to describing the proposed outlier detection method;
the details are presented in \autoref{ssec_3_1} along with a rule for selecting the adjustment factor. \autoref{ssec_3_2} emphasizes the robustness of the overall procedure against outliers.
The proposed methods are useful and outstanding for identifying functional shape outliers,
which can be challenging for the existing depth notions,
as illustrated in the numerical study in \autoref{sec4}.
Data applications are described in \autoref{sec5}, 
indicating that our methods are applicable to some complex structured functional data.
All proofs, additional simulations, and extra results from the real data analyses can be found in the supplement.


\section{Regularized Halfspace Depth} \label{sec2}

In \autoref{ssec_2_1}, we start by defining the regularized halfspace depth for functional data and verifying the non-degeneracy property. 
\autoref{ssec_2_2} then describes and digests the desirable theoretical properties of the RHD, while \autoref{ssec_2_3} is devoted to defining the sample version of the RHD and establishing its consistency.
A practical computational algorithm for approximating the sample RHD based on random projections is provided along with its theoretical justification in \autoref{ssec_2_4}.
Finally, we discuss how to select tuning parameters involved in the computation of the RHD in \autoref{ssec_2_5}.


\subsection{Motivation and Definition} \label{ssec_2_1}


Let $X$ be a random function defined on a probability space $(\Oa, \ff, \pr)$ that takes values in the underlying Hilbert space $\HH$ with inner product $\langle \cdot, \cdot \rangle$. 
We suppose that $\HH$ is infinite-dimensional and separable.
The original \emph{Tukey's halfspace depth} \citep{tukey75} of $x \in \HH$ with respect to the probability measure $P_X$ induced by $X$ is, 	for $x \in \HH$, 
\begin{align}
	D(x) = D(x; P_X) 
	& = \inf_{v \in \HH: \|v\|=1} \pr( \langle X - x, v \rangle \geq 0 ). \label{eqOHD2}
\end{align}
{
However, the direction set $\SSS \equiv \{v \in \HH: \|v\|=1\}$ in the definition of the Tukey's halfspace depth \eqref{eqOHD2} is too rich due to the infinite dimension of $\HH$.
This set $\SSS$ contains an infinite orthonormal set in $\HH$, which results in $\SSS$ not being totally bounded \cite[Chapter~4]{mac09} and the infimum in \eqref{eqOHD2} being too small.
As a result, the naive definition in \eqref{eqOHD2} may become degenerate in an infinite dimensional space \citep{DGC11}, as described next.
}

\begin{thm}[Degeneracy of Tukey's halfspace depth] \label{thmDegeneracy}
	{Suppose that $X$ has independent functional principal component (FPC) scores, that is, $\{\langle X, \phi_j \rangle\}_{j=1}^\infty$ are independent.}
	Then, it holds that $D(x; P_X) = 0$ for $P_X$-almost surely all $x \in \HH$. 
\end{thm}

We overcome this degeneracy issue and define a new notion of halfspace depth by introducing a regularization step. Let $\HH$ be an infinite-dimensional Hilbert space, and define by $x \otimes y:\HH\to\HH$, the tensor product of two elements $x,y \in \HH$, as a bounded linear operator $(x \otimes y)(z) = \langle z,x \rangle y$ for $z \in \HH$. Under a finite second moment assumption $\eo[\|X\|^2]<\infty$, the covariance operator of $X$ is defined as $\ga \equiv \eo[(X - \eo[X]) \otimes (X - \eo[X])]$.
Since the covariance operator $\ga$ is self-adjoint, non-negative definite, and compact, by spectral decomposition, it admits the eigen-decomposition $\ga = \sum_{j=1}^\infty \g_j (\phi_j \otimes \phi_j)$,
where $(\ld_j, \phi_j)$ is the $j$-th eigenvalue--eigenfunction pair of $\ga$ with the properties that $\g_1 \geq \g_2 \geq \cdots \geq 0$, $\g_j \to 0$ as $j\to \infty$, and the eigenfunctions $\{\phi_j\}_{j=1}^\infty$ form a complete orthonormal system for the closure of the image of $\ga$ \citep{HE15}. 
We further assume that the eigenvalues are positive and strictly decreasing, i.e., $\g_1>\g_2>\cdots>0$,
in order to avoid non-interesting cases, such as where random elements supported only on a finite-dimensional subspace of $\HH$.

The proposed regularized halfspace depth is defined next.

\begin{defn}
	The \emph{regularized halfspace depth (RHD)} of $x \in \HH$ with respect to $P_X$ is defined as 
	\begin{align}
		D_\ld(x) = D_\ld(x; P_X)
		= \inf_{v \in \HH: \|v\|=1, \|\ga^{-1/2}v\| \leq \ld} \pr( \langle X-x,v \rangle \geq 0), \label{eqRHD2}
	\end{align}
	where $\ld \in (0, \infty)$ is a \emph{regularization parameter} and $\ga^{-1/2} \equiv \sum_{j=1}^\infty \g_j^{-1/2} (\phi_j \otimes \phi_j)$.
\end{defn}

{
We refer to the norm $\| v \|_{\HH(\ga)} \equiv \|\ga^{-1/2}v\|$ as the reproducing kernel Hilbert space (RKHS) norm because if $\HH=L^2([0,1])$ and $X$ has a continuous covariance function $G(s,t)=\cov(X(s), X(t))$, $s,t\in[0, 1]$, then $\HH(\ga) \equiv \{ v \in L^2([0,1]): \|\ga^{-1/2}v\| < \infty \}$ is an RKHS with reproducing kernel $G$ \cite[cf.][]{wahba73} and an infinite-dimensional dense subspace of $\HH$.
The definitions $\| v \|_{\HH(\ga)}$ and $\HH(\ga)$ are generally valid given any Hilbert space $\HH$.

However, the RKHS $\HH(\ga)$ is still large since
it contains all eigenfunctions, 
leading to extremely small halfspace probabilities and subsequent degeneracy.
For this reason, we regularize the RKHS norm $\| v \|_{\HH(\ga)}$ of projection direction $v$ using a pre-specified number $\ld \in (0,\infty)$.
The resulting set of projection directions used to define the RHD $D_\ld$ is written as
\begin{align} \label{eqProjDir}
	\vv_\ld \equiv \{ v \in \HH: \|v\|=1, \|\ga^{-1/2}v\| \leq \ld \}. 
\end{align}
The set $\vv_\ld$ is then reasonably small in the sense that it is totally bounded (cf.~Proposition~S4 in the supplement).

The regularization is the key difference of our method from other existing depth notions. 
For a given $\ld \in (0,\infty)$, we restrict the set of unit-norm projection directions used in the halfspace probabilities to $\vv_\ld$ in \eqref{eqProjDir}, which requires the RKHS norm of the projections to be at most $\ld$.
This will resolve the degeneracy issue as shown below in Theorem 2. 
To the best of our knowledge, regularization has never been used in the depth literature in the context of either finite or infinite dimensional spaces.

Without regularization, degeneracy occurs for the original Tukey's depth \eqref{eqOHD2} in $\HH$ as explained in \autoref{thmDegeneracy}. 
In contrast, our projection set $\vv_\ld$ from \eqref{eqProjDir} is moderately sized, 
since it contains a collection of infinite linear combinations of the eigenfunctions 
with the constraint that the coefficients of non-leading eigenfunctions must be small.
For instance, let $a_j = [(\gamma_j2^{-j}) /(\sum_k \gamma_k2^{-k})]^{1/2}$ for $j=1,2,\dots$. 
If  $\lambda \ge (\gamma_1/2)^{-1}$, then $\sum_j a_j \phi_j \in \vv_\ld$. 
However, $\phi_j \in \vv_\ld$ if and only if $\gamma_j^{-1/2} \le \lambda$, so $\vv_\ld$ contains only finitely many eigenfunctions; 
see also Proposition~S1 in the supplement. 
This enables/allows our proposed RHD to avoid the degeneracy issue. We will further show in \autoref{ssec_2_2} that RHD is a flexible data depth method with desirable theoretical properties.

\autoref{thmNonDengeracy} shows that the RHD is free of degeneracy. 
	To state the theorem, let $\mu = \eo[X]$ and $F_v$ denote the cumulative distribution of the standardized projection $\langle X-\mu, v \rangle / \|\ga^{1/2} v\|$.
	For each $\ld, t \in (0,\infty)$, we consider the following condition on the tail probabilities of the standardized projections:
	
	\vspace{0.2in}
	
	\noindent
	Condition~$E(\ld, t)$: $\sup_{v \in \vv_\ld} F_v(t) <1$.

	\vspace{0.1in}
}

\begin{thm}[Non-degeneracy of the RHD] \label{thmNonDengeracy}
	{
	Let $\ld \in (0,\infty)$ be fixed.
	If there exists $t \in (0,\infty)$ such that Condition~$E(\ld, t)$ holds, then $D_\ld(x)>0$ for any $x \in \HH$ with $\|x-\mu\|\leq t / \ld$. 
	Furthermore, if Condition~$E(\ld, t)$ holds for each $t \in (0,\infty)$, then the RHD, $D_\ld$, is positive everywhere in $\HH$.
	}
	
		
\end{thm}


{
Condition~$E(\ld, t)$ for \autoref{thmNonDengeracy} ensures that the upper tail probabilities of the standardized projection are not too small across the direction set $\vv_\ld$.
This tail probability condition is mild and general,
which holds especially when the standardized projection distribution $F_v$ does not depend on $v \in \vv_\ld$.
This encompasses a wide range of random elements, particularly those whose standardized projections follow a well-known distribution, those whose FPC scores are dependent, or those whose higher moments are infinite, as listed below.

\begin{eg} \label{eg1}
	
	A variety of random elements can satisfy Condition~$E(\ld, t)$ if the projections follow a well-known distribution. 
	In particular, the projection $\langle X, v \rangle$ could follow distributions such as normal, logistic, or double exponential distributions, which are determined only by location and scale parameters.
	For instance, if we define the random element $X$ such that its projection $\langle X, v \rangle$ follows the logistic distribution, then the standardized projection distribution $F_v$ is independent of $v \in \vv_\ld$ as $F_v (u) = \{1+e^{-u}\}^{-1}$ for each $u \in \R$. Therefore, Condition~$E(\ld, t)$ is satisfied for each $t \in (0,\infty)$. 
	
\end{eg}

\begin{eg} \label{eg2}
	
	Condition~$E(\ld, t)$ can hold even for random elements with dependent FPC scores or with infinite higher moments.
	Suppose $X = \xi R$ where $R$ is the Gaussian element with mean zero and $\xi$ is the standardized \textsf{t}-distribution with degree of freedom $\nu>2$.
	Its FPC scores $\langle X, \phi_j \rangle = \xi \langle R, \phi_j \rangle$ are dependent through the common latent variable $\xi$
	and the higher moments of $X$ might not exist depending on $\nu > 2$. 
	Since the covariances of $X$ and $R$ are equal, 
	the distribution of the projection $\langle X, v \rangle = \xi \langle R, v \rangle$ is given as $F_v(u) = \pr \left( \xi Z \leq u \right)$ for each $u \in \R$, where $Z$ denotes a standard normal variable.
	Consequently, the distribution function $F_v$ does not depend on the direction $v \in \vv_\ld$,  
	and hence, Condition~$E(\ld, t)$ holds for each $t \in (0,\infty)$.
\end{eg}

\begin{rem}
	Condition~$E(\ld, t)$ in \autoref{thmNonDengeracy} allows handling the infinite dimensionality of $\HH$.
	In functional data analysis, it is quite common to consider some distributional conditions similar to Condition~$E(\ld, t)$ for theoretical development 
	as the class of random elements in an infinite dimensional space is redeemed extremely large.
	Common assumptions include finite fourth moments \citep{HH07, CMS07, GK07, GHK10} or finite mixed moments between the FPC scores \citep{YM10, FN22}, with some studies even assuming all moments to be finite \citep{DH10}.
	Other works impose more specific distributional assumptions such as continuity of $X$ \citep{CDJ13}, (centered) elliptical processes \citep{LR23}, or subgaussianity on the vector of the FPC scores \cite{LL21}.
	Compared to these, Condition~$E(\ld, t)$ is indeed a relatively mild assumption that encompasses many distributions applicable to practical problems (cf.~Examples~\ref{eg1}-\ref{eg2}).
	Therefore, incorporating some distributional conditions on $X$ is reasonable to achieve non-degeneracy of the RHD, allowing us to leverage its practical benefits.
\end{rem}
}

%
%

In addition to resolving degeneracy, regularization parameter $\lambda$ also provides flexibility on emphasizing the shape and/or magnitude of the functional observations.
Given a small $\lambda$, the RHD focuses more on the overall magnitude of the input functions since the projection set concentrates around the leading eigenfunctions, which tend to capture lower frequency variations.
On the other hand, as $\lambda$ increases, the RHD increases its emphasis on variation along the higher-frequency directions given by the non-leading eigenfunctions.
This property of the RHD will be utilized later when we develop a new method for shape outlier detection.
{The regularization parameter $\ld \in (0,\infty)$ is a user-specific parameter that depends on the purposes of the analysis. The selection of $\ld$ will be discussed in \autoref{ssec_2_5}.}

{
Surprisingly, the RHD is shown to be robust against outliers 
despite utilizing the covariance, which is generally assumed to exist in functional data literature but typically considered non-robust.
Moreover, the RHD can be made more robust by restricting the directions in $\vv \equiv \{ v \in \HH: \|v\|=1 \}$ to a general subset of $\vv$ that does not require the finite moments.
One option is to substitute the RHKS norm $\|\ga^{-1/2}v\|$ with another RKHS norm with a different reproducing kernel.
However, such general subsets may not reflect the variability in shape and magnitude of the functional observations.
We show that by imposing a second moment condition, the RHD offers numerous practical benefits, such as detecting many types of shape outliers in complex functional data sets; see \autoref{sec4} for numerical performance. 
These advantages are usually not obtained when using a single functional depth such as modified band depth \citep{LR09}, extremal depth \citep{NN16}, integrated depth \citep{NGH17}, and so on;
nonetheless, using a combination of multiple depth functions can give similar performance but adds complexity  \citep{NGH17, DMSG20}. 
Despite the covariance being non-robust, the proposed methods based on RHD exhibit resistance to many types of outliers and are shown to be useful in practice. This has been illustrated and discussed in \autoref{ssec_4_4} and \autoref{ssec_3_2}, respectively. 
}

Our proposed RHD can be constructed whenever the underlying function space is a separable Hilbert space, for example, $L^2 \equiv L^2([0,1]) = \left\{ f:[0,1]\to\R | \int_0^1 f(t)^2 dt < \infty  \right\}$ with inner product $\langle f_1, f_2 \rangle \equiv \int_0^1 f_1(t) f_2(t) dt$ for $f_1,f_2 \in L^2$. More generally, the RHD is also applicable to functional data taking values in a multivariate space \cite[cf.][]{CHSV14, HRS15, LSLG14} 
given an appropriately defined inner product. For solidity, in this work, we focus on univariate curves in $\HH=L^2$.

\subsection{Theoretical Properties} \label{ssec_2_2}

For a depth function to be intuitively sound, it should provide a center-outward ordering of the data and satisfy some desirable properties. 
\cite{Liu90} discussed some properties satisfied by the proposed simplicial depth. \cite{ZS00} formally postulated a list of desirable properties that should be satisfied by a well-designed depth function and showed that the original Tukey's halfspace depth satisfies all these properties. 
\cite{NB16,GN17} among others extended such depth properties for functional data. 
We show some basic properties of the RHD in the following theorems which extend the postulated multivariate depth properties introduced by \cite{ZS00} to the infinite-dimensional Hilbert space $\HH$. The proofs are deferred to Section~S1.4 in the Supplementary Materials.

\begin{thm} \label{thmBasic4}
	\hfill
	\begin{enumerate}[(a)]
		\item (Isometry invariance) 
		Let $A:\HH\to\HH$ is a bounded linear operator and $b \in \HH$. 
		If $A$ is a surjective isometry, 
		then we have $D_\ld(Ax+b, P_{AX+b}) = D_\ld(x, P_X)$ for each $x \in \HH$.
		
		\item (Maximality at center) 			
		Suppose that $X$ is halfspace symmetric about a unique center $\mu \in \HH$ in the sense that $P_X(H) \geq 1/2$ for each $H \in \hh(\mu)$, where $\hh(x)$ denotes the collection of all closed halfspaces containing $x \in \HH$. 
		Then, if $\ld > \g_1^{-1/2}$ where $\gamma_1$ is the largest eigenvalue, we have $D_\ld(\mu) = \sup_{x \in \HH} D_\ld(x)$.

		\item (Monotonicity relative to deepest point) Let $\theta \in \HH$ be such that $D_\ld(\theta) = \sup_{x \in \HH} D_\ld(x)$. Then, it holds for any $x\in\HH$ that $D_\ld(x) \leq D_\ld(\theta + \ap(x-\theta))$, $\ap \in [0,1]$.

		\item (Vanishing at infinity) 
		{
		Let $\uu_\ld \equiv \{ x \in \HH: \|\ga^{-1/2}x\| \leq \ld\|x\| \}$.
		Then, for any sequence $\{x_n\} \subseteq \uu_\ld$ with $\|x_n\| \to \infty$ as $n\to\infty$, we have $D_\ld(x_n) \to 0$ as $n\to\infty$.}
		
	\end{enumerate}
\end{thm}

\autoref{thmBasic4}(a) extends the affine invariance property \cite[cf.][]{ZS00} to an infinite dimensional space and establishes this property for the proposed RHD
because any surjective isometry between two normed spaces is affine by the Mazui--Ulam theorem \citep{vai03}.
Properties (b) and (c) in \autoref{thmBasic4} respectively states that the RHD is maximized at the unique center $\mu$ and it is nonincreasing along any ray leaving from the deepest curve $\mu \in \HH$. These properties are analogous to their Euclidean versions in \cite{ZS00}, but the proof for (b) is more involved and uses properties of the projection set $\vv_\ld$.
{\autoref{thmBasic4}(d) is a weakened version of vanishing at infinity property where we need some restrictions such as the unnormalized direction set $\uu_\ld$ to limit the behavior of the depth function; see Section~S1.4 in the Supplementary Materials for another version of the vanishing at infinity property with the finite-dimensional assumption and a counterexample. }

The next theorem establishes the continuity of the depth function and topological properties of its level sets. 
For $\tau \in [0, \infty)$, we define the \emph{upper level set} $G_{\tau, \ld}$ of $D_\ld$ at level $\tau$ as
$G_{\tau, \ld} \equiv \{ x \in \HH: D_\ld(x) \geq \tau \}.$
The following theorem extends analogous properties for the finite-dimensional data case shown in Theorem~2.11 of \cite{ZS00}. 

\begin{thm} \label{thmLevelSets}
	\hfill
	\begin{enumerate}[(a)]
		\item 
		The depth function $D_\ld:\HH\to[0, \infty)$ is upper semi-continuous.
		
		\item 
		The upper level sets $\fgg_\ld \equiv \{G_{\tau, \ld}\}_{\tau \in [0, \infty)}$ are nested, closed (hence, complete), and convex in $\HH$.
		
		\item
		Let $\V \equiv \mathrm{span} \{ \phi_1, \dots, \phi_m \} \subseteq \HH$ be a finite-dimensional subspace spanned by the first $m$ eigenfunctions. 
		Then, for any $\tau$, $G_{\tau, \ld} \cap \V$ is compact. 
	\end{enumerate}
\end{thm}

The upper semi-continuity of the RHD $D_\ld$ and the closedness (and hence, the completeness) of the level set $G_{\tau, \ld}$ are closely related.
The compactness follows from vanishing at infinity and the restriction onto a finite dimensional subspace of $\HH$. 
Without this restriction, the compactness of the level sets may not be achieved as described in Section~S1.4 of the Supplementary Materials since vanishing at infinity fails.

\subsection{Sample RHD and its Consistency}	\label{ssec_2_3}

The population version of the RHD from \eqref{eqRHD2} cannot be computed in practice. To define its sample version estimated from the observed functions, let $X_1, \dots, X_n$ be independently and identically distributed (iid) copies of $X$ with empirical distribution $\hat{P}_n$ defined as $\hat{P}_n(B) \equiv n^{-1} \sum_{i=1}^n \I(X_i \in B)$ for a Borel set $B \in \bb(\HH)$, where $\I$ denotes the indicator function.. 
The sample covariance operator $\hat{\ga}_n$ is defined as $\hat{\ga}_n \equiv n^{-1} \sum_{i=1}^n (X_i - \bar{X}) \otimes (X_i - \bar{X})$ with sample mean $\bar{X} \equiv n^{-1} \sum_{i=1}^n X_i$. 
Write the spectral decomposition of the operator $\hat{\ga}_n$ as $\hat{\ga}_n = \sum_{j=1}^n \hat{\g}_j (\hat{\phi}_j \otimes \hat{\phi}_j)$,
where $(\hat{\ld}_j, \hat{\phi}_j)$ is the $j$-th eigenpair of $\hat{\ga}_n$ for $j=1,\dots,n$, with $\hat{\g}_1 \geq \cdots \geq \hat{\g}_n \geq 0$  \citep{HE15}.

To define the sample version of the RHD, the set of projection directions is obtained as
\begin{align} \label{eqProjDirSample}
	\hat{\vv}_{\ld, J}  \equiv \{ v \in \mathrm{span} \{ \hat{\phi}_1, \dots, \hat{\phi}_J \}: \|v\|=1, \|\hat{\ga}_J^{-1/2} v\| \leq \ld  \},
\end{align}
where $\hat{\ga}_J^{-1/2} \equiv \sum_{j=1}^J \hat{\g}_j^{-1/2} \hat{\pi}_j$ is a $J$-truncated inverse of the square root operator $\hat{\ga}_n^{1/2} = \sum_{j=1}^n \hat{\g}_j^{1/2} \hat{\pi}_j$.
Here, the truncation parameter $J = J(n)$ may depend on the sample size $n$ with $J(n) \to \infty$ as $n\to\infty$.
We define the \emph{sample RHD} of $x \in \HH$ (with respect to $\hat{P}_n$) as
\begin{align}
	\hat{D}_{\ld,n}(x) = \hat{D}_{\ld,n}(x, \hat{P}_n)
	= \inf_{v \in \hat{\vv}_{\ld, J}} n^{-1} \sum_{i=1}^n \I( \langle X_i - x, v \rangle \geq 0). \label{eqRHDsample2}
\end{align}

To establish the uniform consistency of the sample RHD $\hat{D}_{\ld, n}(x)$ for the population RHD $D_\ld(x)$, 
we need the following continuity/anti-concentration assumption imposed on the distribution of the projection of $X$:
for a regularization parameter $\ld \in (0,\infty)$ and a totally bounded subset $\B \subseteq \HH$,
\vspace{0.2in}

\noindent
Condition~$C(\ld, \B)$: $\dsp \lim_{r \to 0} \limsup_{k \to \infty} \sup_{x \in \B, v \in \vv_\ld} \pr( | \langle X - x, \Pi_k v \rangle | \leq r) = 0$,

\vspace{0.2in}

\noindent
where $\Pi_k \equiv \sum_{l=1}^k (\phi_k \otimes \phi_k)$ denotes the projection operator onto the linear space spanned by the first $k$ eigenfunctions for integer $k \geq 1$.	

This condition is satisfied by a large class of random functions $X$ where the projection scores $ \langle X, \phi_j \rangle$, $j=1, 2, \dots$ are independent and continuous; in particular, Gaussian random functions. 

\begin{eg}
	If $X$ is Gaussian, for each $v \in \vv_\ld$ and $x \in \B$, 
	a truncated projection $\langle X - x, \Pi_k v \rangle$ follows the normal distribution with mean $\langle \mu - x, \Pi_k v \rangle$ and variance $\|\ga^{1/2} \Pi_k v\|^2$
	Here, $\Pi_k v \neq 0$ for all $\vv_\ld$ and for sufficiently large $k$ by Lemma~S1 in the supplement.
	Moreover, the mean and standard deviation are bounded by Lemma~S2
	as $\sup_{x \in \B, v \in \vv_\ld} |\langle \mu - x, \Pi_k v \rangle| \leq \|\mu\| + \sup_{x \in \B} \|x\|<\infty$ and $\liminf_{k\to\infty} \inf_{v \in \vv_\ld} \|\ga^{1/2} \Pi_k v \| \geq (2\ld)^{-1} > 0.$
	By the properties of normal distribution,
	$$\limsup_{k\to\infty} \sup_{x \in \B, v \in \vv_\ld} \pr( | \langle X - x, \Pi_k v \rangle | \leq \tau)
	\leq \Phi(2\ld\tau) - \Phi(-2\ld\tau) \to 0$$
	as $\tau \to 0$.
	Hence, Condition~$C(\ld, \B)$ holds for Gaussian element $X$.
\end{eg}

To verify the consistency theorem, in addition to Condition~$C(\ld, \B)$, we impose some standard regularity conditions on $X$, denoted as Conditions~(A1)-(A4) in Section~S1.5.3 of the supplement.
Similar or stronger conditions are commonly imposed in the functional principal component analysis literature for consistent estimation of the eigen-components \citep{CMS07, HH07}.
These conditions are applied in the proofs involving perturbation theory for functional data (cf.~Sections~S1.5.3-S1.5.4 of the supplement).
The consistency result is established in the following theorem.

\begin{thm}  \label{thmConsistency}
	
	Let $\ld \in (0,\infty)$ be fixed and $\B \subseteq \HH$ be a totally bounded subset.
	Suppose that 
	Condition~$C(\ld+\ld_0, \B)$ holds for some $\ld_0 \in (0,\infty)$,
	Conditions~(A1)-(A4) in the supplement hold, 
	and $n^{-1/2} J^{7/2} (\log J)^2 = O(1)$ as $n\to\infty$.
	Then, 
	as $n\to\infty$, we have  
	\begin{align*}
		\sup_{x \in \B} |\hat{D}_{\ld, n}(x) - D_\ld(x)| \xrightarrow{\pr} 0.
	\end{align*}
\end{thm}


{
The proof for this result is included in the supplement. 
A main component of the proof is to establish the Glivenko--Cantelli theorem for the empirical process $\{\hat{P}_n(H_{x,v}): v \in \vv_\ld, x \in \B\}$ indexed by an infinite-dimensional class of halfspaces, 
where $H_{x,v} \equiv \{ y \in \HH: \langle y-x,v \rangle \geq 0 \}$ denote the closed halfspace with normal vector $v \in \HH$ containing $x \in \HH$ on its boundary.
Empirical process results for infinite-dimensional space are challenging to show 
except for the results on bracketing numbers of the classes of either smooth or monotone functions by \citet[Section~2.7]{VW96} and
Glivenko--Cantelli and Donsker theorems over some finite dimensional subspace by \cite{CC14b}.
Developing empirical process theory for functional data may not be feasible without such restriction on the class of functions to reasonably small subclasses.
The main reason is that, in the infinite-dimensional Hilbert space $\HH$, sufficiently small balls (e.g., balls with radius smaller than $\sqrt{2}$) fail to cover the closed unit ball;
this implies the closed unit ball in $\HH$ is not only non-totally-bounded \cite[e.g.,][]{mac09, HE15}, but it may also not have a finite covering number.
As a related discussion, we refer to \cite{Cara00}, showing the closed unit ball in an infinite dimensional Banach space is not a Glivenko-Cantelli class.
Here, we consider a totally bounded subset $\B$ for such restriction.
The proof consists of showing the total boundedness of $\vv_\ld$ and asymptotic uniform equicontinuity of the stochastic process $\{\hat{P}_n(H_{x,v}): v \in \vv_\ld, x \in \B\}$ (cf.~Section~S1.4.2 of the supplement), which is more involved than directly applying the covering or bracketing numbers \cite[cf.][Chapter~2]{VW96}.
Please refer to the respective paragraphs following Theorems~5-6.
}

\subsection{Practical Computation} \label{ssec_2_4}

The sample RHD involves an infimum over the projection set $\hat{\vv}_{\ld, J}$ which makes the computation hard; even for the original Tukey's depth in $\R^d$ the exact computation can be prohibitive for dimension $d \ge 4$ \citep{DM16}. 
To approximate the sample RHD, we adopt a random projection approach.

For $\hat{v} = \sum_{j=1}^J \hat{a}_j \hat{\phi}_j \in \mathrm{span}\{\hat{\phi}_1, \dots, \hat{\phi}_J\}$ with $\hat{\bm{a}} = (\hat{a}_1, \dots, \hat{a}_J)^\top \in \R^J$, 
it holds that 
\begin{align*}
	\|\hat{v}\|^2 = \sum_{j=1}^J \hat{a}_j^2 = \|\hat{\bm{a}}\|_{\R^J}^2
	\quad\text{and}\quad
	\|\hat{\ga}_J^{-1/2}\hat{v}\|^2 = \sum_{j=1}^J \hat{\g}_j^{-1} \hat{a}_j^2 = \| \hat{\rr}_J^{-1/2} \hat{\bm{a}} \|_{\R^J}^2
\end{align*}
where $\hat{\rr}_J \equiv \mathrm{diag}(\hat{\g}_1, \dots, \hat{\g}_J)$. Then, $\hat{\vv}_{\ld, J}$ can be written as 
\begin{align*}
	\hat{\vv}_{\ld, J} 
	= \left\{ \hat{v} = \sum_{j=1}^J \hat{a}_j \hat{\phi}_j \in \mathrm{span}\{\hat{\phi}_1, \dots, \hat{\phi}_J\}: \hat{\bm{a}} = (\hat{a}_1, \dots, \hat{a}_J)^\top \in \hat{\aaa}_{\ld, J} \right\}
\end{align*}
where $\hat{\aaa}_{\ld, J} \equiv \{ \hat{\bm{a}} \in \R^J: \|\hat{\bm{a}}\|_{\R^J}=1, \| \hat{\rr}_J^{-1/2} \hat{\bm{a}} \|_{\R^J} \leq \ld \}$.
Note that the set $\hat{\aaa}_{\ld, J}$ is the intersection of the (surface of) unit sphere and a (solid) ellipsoidal region centered at the origin in $\R^J$.
Then, the sample RHD is written as
\begin{align}
	\hat{D}_{\ld, n}(x) 
	= \inf_{\hat{\bm{a}} \in \hat{\aaa}_{\ld, J}} n^{-1} \sum_{i=1}^n \I((\hat{\bm{X}}_i - \hat{\bm{x}})^\top \hat{\bm{a}} \geq 0) \label{eqRHDsample3_vector}
\end{align}
where $\hat{\bm{X}}_i = [\langle X_i, \hat{\phi}_j \rangle]_{1 \leq j \leq J}$ and $\hat{\bm{x}} = [ \langle x, \hat{\phi}_j \rangle ]_{1 \leq j \leq J}$ are $J$-dimensional vectors in $\R^J$.
Although \eqref{eqRHDsample3_vector} is reminiscent of sample Tukey's depth applied on the data $\hat{\bm{X}}_i$, they are remarkably different in their designs.
The RHD targets infinite-dimensional data and allows for an increasing $J$. 
The depth values are insensitive to increase in $J$ when it is moderately large ($J\ge 6$ as demonstrated in our numerical studies) thanks to the regularization set which stabilizes the depth.
On the other hand, Tukey's depth is most often applied in a fixed low dimension; when the dimension increases, 
Tukey's depth become less capable of distinguishing points 
because of the concentration of measure and the degeneracy issue.

To approximate the infimum, we draw $M$ random vectors $\hat{\bm{a}}_m = (\hat{a}_{m1}, \dots, \hat{a}_{mJ})^\top \iid \nu$ for $m=1,\dots,M$ from some continuous distribution $\nu$ supported on $\hat{\aaa}_{\ld, J} \subseteq \R^J$.
The projection set is approximated by the finite set
\begin{align}
	\tilde{\vv}_{\ld, J, M} \equiv \left\{ \tilde{v} = \sum_{j=1}^J \hat{a}_{mj} \hat{\phi}_j : m=1, \dots, M \right\}. \label{eqVTilde}
\end{align}
Then, the sample depth $\hat{D}_{\ld,n}(x) = \hat{D}_{\ld,n}(x, \hat{P}_n)$ in \eqref{eqRHDsample3_vector} is approximated by
\begin{align}
	\tilde{D}_{\ld, n, M}(x) 
	= \tilde{D}_{\ld, n, M}(x, \hat{P}_n, \{\hat{\bm{a}}_m\}_{m=1}^M) 
	= \min_{m=1, \dots, M} n^{-1} \sum_{i=1}^n \I((\hat{\bm{X}}_i - \hat{\bm{x}})^\top \hat{\bm{a}}_m \geq 0). \label{eqRHDapprox2}
\end{align}
Rejection sampling is applied to draw the projection directions $\hat{\bm{a}}_m$. 
To increase acceptance in the target region $\hat{\aaa}_{\ld, J}$, we set the proposal distribution of projection directions $\hat{\bm{a}}_m$ to that of $ \bm{z}/\|\bm{z}\|_{\R^J}$ where $\bm{z} \sim \nd(\bm{0}, \hat{\rr}_J)$, 
a non-isotropic distribution on the unit sphere.

The following theorem justifies the approximated sample RHD in \eqref{eqRHDapprox2} for a sufficiently large $M$, for which 
the proof is given in Section~S1.6 in the supplement.
\begin{thm} \label{thmApproxSampleDepth}
	Let $\ld \in (0,\infty)$ be fixed and $\B \subseteq \HH$ be a totally bounded subset.
	Suppose that the assumptions in \autoref{thmConsistency} hold.
	Then, for each $\e>0$, the following event occurs with probability tending to 1 as $n\to\infty$: for a continuous probability distribution $\nu$ supported on $\hat{\aaa}_{\ld, J}$ 
	and a sequence $\{\hat{\bm{a}}_m\}_{m=1}^\infty$ of iid random vectors drawn from $\nu$, as $M\to\infty$,
	\begin{align*}
		\nu \left( \sup_{x \in \B} \left| \tilde{D}_{\ld, n, M}(x) - \hat{D}_{\ld,n} (x) \right| > \e \right) \to 0.
	\end{align*}
\end{thm}

The key point of the proof is to show a type of asymptotic equicontinuity of a process $\{ \hat{P}_n(H_{x,\hat{v}}): x \in \B, \hat{v} \in \hat{\vv}_{\ld, J} \}$,
where the index set also depends on the observed random functions $\xx_n \equiv \{X_i\}_{i=1}^n$. 
Such processes with a changing index set following the sample size $n$ have rarely been studied.
We derived this theorem by constructing a correspondence from $\hat{\vv}_{\ld, J}$ to $\vv_\ld$, relating this correspondence with the asymptotic equicontinuity of the original process $\{ \hat{P}_n(H_{x,v}): x \in \B, v \in \vv_\ld \}$, and using the total boundedness of $\vv_\ld$.

The population \eqref{eqRHD2} and sample RHD \eqref{eqRHDsample2} are monotonically decreasing as $\lambda$ increases.
In order for the approximate RHD to enjoy the same property, namely $\lambda \mapsto \tilde{D}_{\ld, n, M}(x)$ is decreasing in $\lambda$ when all other quantities are fixed, we employ the following procedure.
Let $\Ld \equiv \{\ld_1, \dots, \ld_Q\}$ be a finite set of regularization parameters that we will use to evaluate the approximate RHD. 
We draw a single set of random directions $\bm{z}_1, \dots, \bm{z}_M \sim \nd(0, \hat{\rr}_J)$ and set $\hat{\bm{a}}_m = \bm{z}_m/\|\bm{z}_m\|_{\R^J}$ as explained above, and use the directions $\{\hat{\bm{a}}_m\}_{m=1}^M$ to compute all depth values $\{\tilde{D}_{\ld, n, M}: \ld \in \Ld\}$. The monotonicity then follows from  the definition of $\tilde{D}_{\ld, n, M}$.

In practice, instead of specifying $\ld$ for regularization, 
we use the $u$-quantile of the RKHS norms $\nn_M \equiv \{ \| \hat{\rr}_J^{-1/2} \hat{\bm{a}}_m \|_{\R^J} \}_{m=1}^M$ as $\ld$,
where $u \in (0,1)$ is called the \emph{quantile level} for regularization.
The quantile level approach to determine $\lambda$ has several advantages; it depends on the data and avoids choosing values of $\ld$ outsides of the range of the RKHS norms $\nn_M$. 
In our numerical studies, we investigate the performance of RHD using different $u$-quantiles of $\nn_M$ as regularization parameter $\ld$.

\autoref{alg1} summarizes the approximate computation. 
\begin{algorithm2e}[h!]
	\small
	\caption{
		Approximate sample RHD 
	}
	\label{alg1}
	\KwData{
		Random sample $\xx$ and depth evaluation point $x$
	} \nlnonumber
	\KwTune{
		Either regularization parameter set $\Ld$ or quantile level set $\uu$. 
		Also truncation parameter $J$
	}
	
	\KwResult{
		Approximate sample RHD $\tilde{D}_{\ld, n, M}(x)$ for $\ld \in \Ld$
	}
	\For{$m = 1,\dots, M'$}{
		Generate $\bm{z}_m \sim \nd(0, \hat{\rr}_J)$ and compute $\hat{\bm{a}}_m = \bm{z}_m/\|\bm{z}_m\|_{\R_J}$
		
	}	
	
	\If{
		$\Ld$ \emph{is unspecified}
	}{
		\For{$u \in \uu$}{
			Add the $u$-quantile of $\{ \| \hat{\rr}_J^{-1/2} \hat{\bm{a}}_m \|_{\R^J} \}_{m=1}^{M'}$ to $\Ld$
		}
	}

	\For{ $\ld \in \Ld$}{
		
		$\tilde{\aaa}_{\ld, J, M} \leftarrow \emptyset$
		
		\For{$m = 1,\dots,M'$}{
			\If{$\| \hat{\rr}_J^{-1/2} \hat{\bm{a}}_m \|_{\R^J} \leq \ld$}{
				Add $\hat{\bm{a}}_m$ to $\tilde{\aaa}_{\ld, J, M}$
			}\If{$|\tilde{\aaa}_{\ld, J, M}|=M$}{}
		}
		
		$\tilde{D}_{\ld, n, M}(x) \leftarrow \min_{\hat{\bm{a}} \in \tilde{\aaa}_{\ld, J, M}} n^{-1} \sum_{i=1}^n \I((\hat{\bm{X}}_i - \hat{\bm{x}})^\top \hat{\bm{a}} \geq 0)$;
		
	}
	
\end{algorithm2e}

{

\begin{rem}
	The novelty in the RHD arises from the use of the regularization parameter $\ld \in (0,\infty)$.
	Although the approximation of the RHD in \autoref{alg1} and the naive application of the classical halfspace depth to the first $J$ FPC scores appear similar, they are clearly differentiated by the use of $\ld$.
	In our work, the FPC scores serve as a practical method for computing the sample RHD. 
	However, the ``FPCA depth'' without regularization will suffer from degeneracy
	as shown in \autoref{figSim1} and Figure~S1 of the supplement. 
	See Remark~S1 of the supplement for a relevant discussion.
\end{rem}

\begin{rem} \label{remRHDapprox}
	
	The geometry of our projection set $\hat{\vv}_{\ld,J}$, as defined in \eqref{eqProjDirSample}, differs significantly from the classical finite-dimensional case $\SSS_J \equiv \{ a \in \R^J:\|a\|_{\R^J}=1 \}$.
	Specifically, the direction set $\hat{\vv}_{\ld,J}$ is an intersection of an ellipsoid and a sphere, 
	which can be considerably smaller than the classical case $\SSS_J$,
	the entire $J$-dimensional hypersphere.
	As $J$ increases, our sample direction set $\hat{\vv}_{\ld,J}$ targets a constant set $\vv_\ld$ in \eqref{eqProjDir}, ensuring the non-degeneracy of the proposed RHD (cf.~\autoref{thmNonDengeracy}).
	On the other hand, the hypersphere $\SSS_J$ in $\R^J$ used for the finite-dimensional halfspace depth expands to the non-totally-bounded set $\SSS \equiv \{v \in \HH: \|v\|=1\}$ in $\HH$,
    leading to the degeneracy of the original halfspace depth without regularization (cf.~\autoref{thmDegeneracy}).
	As a result, the approximation of the sample RHD by random projections, as justified in \autoref{thmApproxSampleDepth} and described in \autoref{alg1}, is reasonable,
	whereas the classical random projection approach \cite[e.g.,][]{CN08} struggles in higher dimensions $J$.
	Our simulation results indicate that more than $M=1000$ projections generally provides a sufficient approximation for RHD and perform well in various scenarios as seen in Section~4.
\end{rem}
}


\subsection{Selecting Tuning Parameters} \label{ssec_2_5}

The approximate sample RHD $\tilde{D}_{\ld, n, M}$ \eqref{eqRHDapprox2} involves three tuning parameters: the regularization parameter $\ld$ (or equivalently, the quantile level $u$), the truncation level $J$, and the number $M$ of random projection directions.
{
The choice of the most influential parameter $\ld \in (0,\infty)$ depends on the user's analytic goals,
as each $\ld$ can provide different perspectives (e.g., magnitude versus shape) and the optimal $\ld$ for one analysis (e.g., a rank test) may not be ideal for another  (e.g., a classification method).
Moreover, further insight about the data can be obtained by investigating a family of the RHDs $\{D_\ld\}_{\ld \in \Ld}$ with multiple regularization parameters $\Ld \subseteq (0,\infty)$.
The less sensitive tuning parameter $J$ should be a sufficiently large integer,
which may be determined by a common criterion such as the fraction of variance explained (FVE) \cite[cf.][Section~12.2]{KR17}. 
Lastly, a reasonably large value for the final parameter $M$ ensures a sufficient approximation, as supported by \autoref{remRHDapprox}.
We elaborate more details on the selection of these tuning parameters below.
}

The most important parameter, by far, is the regularization $\ld$ since it determines how sensitive the depth is to different modes of variation in the sample, which is a useful feature for analysis.
When $\ld$ is small, projections are constrained to have large components only along the leading eigenfunctions,  
in which case the RHD emphasizes the overall magnitude rather than shape of the curves.
In contrast, when $\ld$ is large, the RHD considers projections along both leading and later eigenfunctions, so it increases its emphasis in shape relative to magnitude. 
This flexibility of the RHD is used and highlighted in our outlier detection method proposed in \autoref{sec3} and is demonstrated numerically in \autoref{ssec_4_3}.

The choice of $\ld$ may depend on the structure of a given dataset and the purpose of the study. 
If a dataset can be explained by a few modes of variation (say, mostly in terms of a vertical shift), a small regularization parameter can be chosen; otherwise a larger regularization parameter should be considered.	
If a class label response exists, the selection of $\ld$ can be tied to depth-based classification performance \citep{GC05,LR06,HRS17}.
Parameter $\ld$ could also be chosen by maximizing the power of RHD-based rank test \citep[for general depth-based rank test, see][]{LS93, LR09} if training samples are available.
In general, practitioners could inspect a range of $\ld$ values and obtain different rankings and outlyingness from the depth values.
This would help understand the dataset from different perspectives in terms of magnitude vs shape.
For practical uses, we recommend as default to set $\lambda$ according to a quantile level $u=0.95$ of the RKHS norms $\nn_M$,
which is implemented in our R package. 
This choice tends to display extremeness in both magnitude and shape while avoiding degeneracy as shown in our numerical studies.

\begin{figure}[b!]
	\centering
	\includegraphics[width=0.99\linewidth]{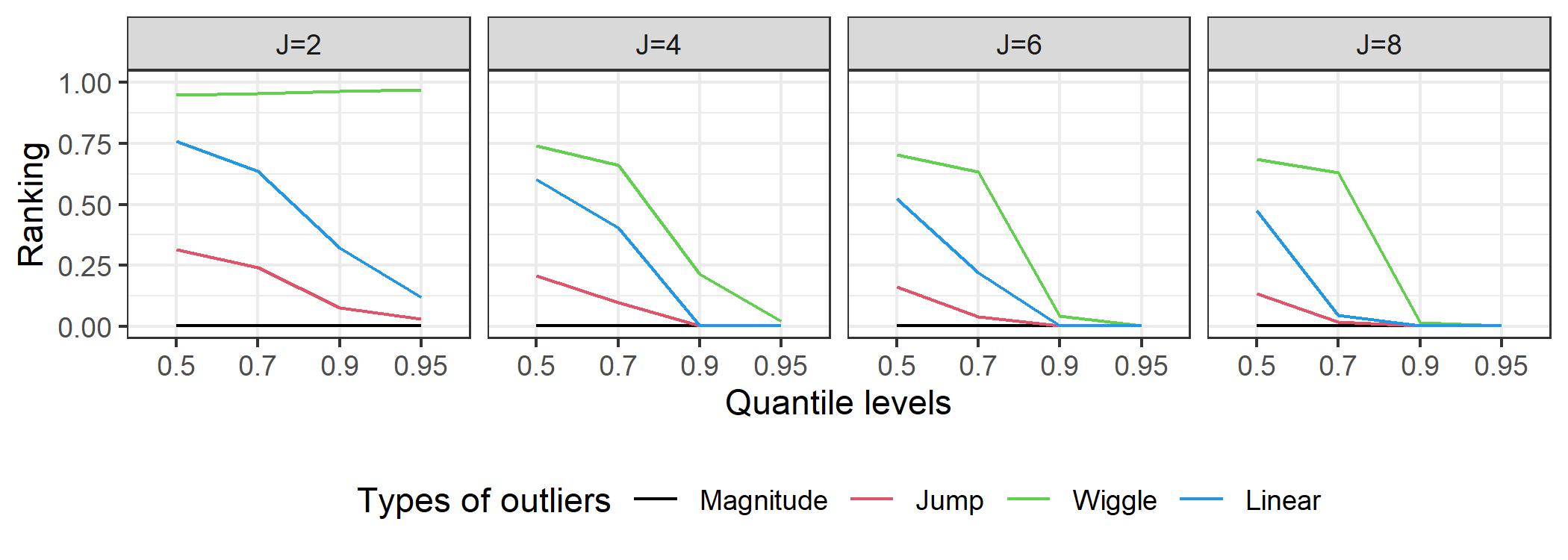}
	\caption{
		RHD-based normalized rankings of outlying curves selected among the ones described in \autoref{ssec_4_1} when $n=400$.
		The least deepest curves have $1/n \approx 0$ normalized rank.
		The results show different truncation levels $J \in \{2,4,6,8\}$ (panels) and four regularization parameters corresponding to the quantile levels $u \in \{0.5, 0.7, 0.9, 0.95\}$ (x-axis).
	}
	\label{figSim2_J_less_sensitive}
\end{figure}

The selection of the truncation level $J$ is a secondary consideration.
It should be chosen as a large enough value, {e.g., based on the FVE,} because $J$ is needed for approximating the population RKHS norm $\|\ga^{-1/2}v\|$ by the sample version $\|\hat{\ga}_J^{-1/2}v\|$.
In \autoref{figSim2_J_less_sensitive}, 
we show the rankings of several types of outliers 
when using the approximate RHD based on different $\lambda$ and $J$ parameters 
to illustrate how sensitive the rankings are to the choice of these parameters. 
A simulated dataset containing non-Gaussian functional data is considered 
where one outlier is added at a time.

A magnitude and three shape outliers of different types were ranked, where the $x$-axis and different panels display varying $\lambda$ and $J$, respectively (see~\autoref{ssec_4_1} for more details).
We define the RHD-based (normalized) rank of a functional observation $X_i$ in the sample $\{X_i\}_{i=1}^n$
as $\tilde{R}_i/n$, 
where $\tilde{R}_i$ denotes the rank of $\tilde{D}_{\ld, n, M}(X_i)$ in $\{\tilde{D}_{\ld, n, M}(X_i)\}_{i=1}^n$,
and therefore, higher ranks correspond to larger depth.
Minimum ranks are assigned to curves with the same depths.
We observe that the results are highly similar 
when $J$ is moderately large, for example, $J \geq 6$ in this case.
In contrast, the rankings for the smallest and largest regularization $\lambda$ are markedly different;
the magnitude outlier gets the lowest rank for all $\lambda$s 
but this is not the case for the shape outliers.
The shape outliers are only detected with higher $\lambda$s
because larger $\ld$ emphasizes shape more strongly.

The least consequential parameter is $M$, which should be chosen to be as large as computation affords. In practice, for example, $M=1000$ is often enough to get a reasonable approximation (cf.~\autoref{remRHDapprox}).

In sum, we recommend to choose a large $M$, a sufficiently large $J$ that may depend on the sample size $n$, and an appropriate $\ld$ for the purposes of the analysis.


\section{Outlier Detection} \label{sec3}

In this section, we describe an outlier detection method developed from the proposed RHD, 
which detects shape as well as magnitude outliers. 
Since having a low depth value is not sufficient to be considered an outlier,
we introduce a novel outlier detection method based on projection directions in \eqref{eqVTilde}.
The overall idea is to first prescreen potential outliers using the RHD, 
and then detect outliers by applying univariate boxplots to data projected onto restricted directions that maximally separate inliers and outliers.
Comprehensive numerical studies in \autoref{ssec_4_3} and the supplement show that our new outlier detection method works well and detects outliers with different complicated shapes.

\subsection{Outlier detection procedure based on the RHD} \label{ssec_3_1}

The detailed outlier detection procedure is described as follows.
We start by defining the most extreme functions with the least (approximate) RHD values as outlier candidates, for which the index set is denoted as
\begin{align*}
	\ii_{\min, \ld} = \ii_{\min, \ld}(\xx_n) \equiv \argmin_{1 \leq i \leq n} \tilde{D}_{\ld, n, M}(X_i).
\end{align*}
Let $X_{i_0}$, $i_0\in \ii_{\min, \ld}$ be an outlier candidate in the sample $\xx_n \equiv \{X_i\}_{i=1}^n$. 
To construct univariate boxplots, we keep track of only the most extreme projection directions for evaluating the halfspace probabilities at $x=X_{i_0}$ for outlier detection.
Among directions in $\tilde{\vv}_{\ld, J, M}$ \eqref{eqVTilde},
suppose that there are $K = K_{i_0}$ directions, denoted as $\{v_{m_k}\}_{k=1}^K$, that minimize the halfspace probability when computing $\tilde{D}_{\ld, n, M}(X_{i_0})$ in \eqref{eqRHDapprox2}, 
that is, 
\begin{align*}
	\{v_{m_k}\}_{k=1}^K 
	\subset \argmin_{1 \leq m \leq M}  \hat{P}_n(H_{X_{i_0}, v_m}).
\end{align*}
For each direction $v_{m_k}$, we obtain projections $\{\langle X_i, v_{m_k} \rangle \}_{i=1}^n$ and label outliers using a univariate boxplot with a given factor $f$ to construct the fence.
The fence is constructed as $[Q_1 - f \times IQR, Q_3+f \times IQR]$ where $Q_1$, $Q_3$, and IQR denoting the first quartile, third quartile, and the interquartile range of the projections, respectively.
We label as outliers $\oo_{i_0, k}$ the functional observations with projection lying outside of the fence along this direction.
The final set of all outliers is obtained as 
\begin{align*}
	\oo_\ld \equiv \left( \bigcup_{i_0 \in \ii_{\min, \ld}} \bigcup_{k=1}^{K_{i_0}} \oo_{i_0, k} \right) \cap \ii_{\min, \ld}.
\end{align*}
The algorithm for outlier detection is given in \autoref{alg2}.

\begin{algorithm2e}[h!]
	\small
	\caption{
		Detect outliers in $\xx_n$ based on the RHD
	}
	\label{alg2}
	\KwData{
		Random sample $\xx_n$ 
	}\nlnonumber
	\KwTune{
		Regularization $\ld$, truncation $J$, and adjustment factor $f$
	}
	
	\KwResult{
		Indices $\oo_\ld$ of outliers in $\xx_n$
	}
	
	Obtain $\tilde{D}_{\ld, n, M}(x), x \in \xx_n$ by invoking \autoref{alg1}
	
	$\ii_{\min, \ld} \leftarrow \argmin_{1 \leq i \leq n} \tilde{D}_{\ld, n, M}(X_i)$
	
	$\oo_\ld \leftarrow \emptyset$
	
	\For{
		$i_0 \in \ii_{\min, \ld}$ and  $\hat{\bm{a}} \in \tilde{\aaa}_{\ld, J}$ 
	}{
		$x \leftarrow X_{i_0}$, $\hat{\bm{X}}_i \leftarrow [\langle X_i, \hat{\phi}_j \rangle]_{1 \leq j \leq J}$, and $\hat{\bm{x}} \leftarrow [\langle x,\hat\phi_j \rangle ]_{j=1}^J$
		
		\If{$n^{-1} \sum_{i=1}^n \I((\hat{\bm{X}}_i - \hat{\bm{x}})^\top \hat{\bm{a}} \geq 0) = \tilde{D}_{\ld, n, M}(x)$}{  
			Form projections $\{ \langle \hat{\bm{X}}_i, \hat{\bm{a}} \rangle \}_{i=1}^n$  and obtain their first quartile, third quartile, and interquartile range as $Q_1$, $Q_3$, and $IQR$
			
			\For{
				$i = 1,\dots, n$
			}{
				\If{
					$\langle \hat{\bm{X}}_i, \hat{\bm{a}} \rangle \notin [Q_1 - f \cdot IQR, Q_3 + f \cdot IQR]$ 
				}{
					$\oo_\ld \leftarrow \oo_\ld \cup \{i\}$
				}
			}
		}
	}
	$\oo_\ld \leftarrow \oo_\ld \cap \ii_{\min, \ld}$;
\end{algorithm2e}

We determine the adjustment factor $f$ in our outlier detection method by a simulation-based method, following the approach in \cite{SG12}. 
The factor $f$ in the functional boxplot is calibrated to the value that assigns 0.7\% observations as outliers in simulated datasets containing only inliers.
The simulated observations follow a Gaussian distribution matching the (empirical) mean and covariance functions of the actual input dataset.
See Algorithm~S1 in the supplement.

\begin{rem}
	This paper does not address the case of clustered functional observations that arise from mixture distributions. Such scenarios might need additional tools such as local depths \citep{PV13} or depth-based classification/clustering methods \citep{LCL12}. However, these problems are beyond the scope of our paper, and we defer to them for future directions.
\end{rem}

\subsection{Robustness of the outlier detection} \label{ssec_3_2}

With a broader interpretation of robustness, the proposed RHD is robust in the following sense. 
First, outliers, even very extreme ones, will not break down our RHD-based outlier detection method as shown in our simulation study (cf.~\autoref{sec4}).
This might sound surprising because our procedure relies on the covariance operator $\ga$ which is non-robust, 
but the sensitivity to outliers actually helps reveal shape outliers 
by capturing the directions where these outliers lie.
Second, the choice of the adjustment factor $f$ by Algorithm~S1 in the supplement is shown to be resistant to outliers, as discussed in Section~S4 of the supplement,
even though the procedure depends on the sample covariance operator $\hat{\ga}_n$ which is (again) non-robust.
This indicates that the whole procedure of our outlier detection method is robust even against very extreme outliers and useful for detecting all types of outliers under our consideration.


\section{Numerical Studies} \label{sec4}

\autoref{ssec_4_1} describes the design of the simulation study along with different types of magnitude and shape outliers under consideration. The performances of the RHD rankings and the proposed outlier detection method are respectively evaluated in \autoref{ssec_4_2} and \autoref{ssec_4_3}.
{Finally, \autoref{ssec_4_4} devotes to study the robustness of the RHD-based median.}

The performance of the RHD is evaluated based on (i) the depth ranks of the simulated outliers in \autoref{ssec_4_2}, (ii) the proportions of falsely/correctly detected outliers in \autoref{ssec_4_3}, {and (iii) the mean squared errors (MSEs) of the RHD-based medians in \autoref{ssec_4_4}}.
For the first study, we compared RHD with other depth notions including MBD \citep{LR09}, extremal depth (ED) \citep{NN16}, as well as the $k$-th order integrated depth by \cite{NGH17}, denoted as $\mathrm{ID}_k$ with order $k \in \{1,2,3\}$.
In the second study, we compared our proposed outlier detection method against functional boxplot \citep{SG11} constructed based on either the MBD or the ED, and with the outlier detection method introduced in \cite{NGH17}.
These methods are implemented in R packages \texttt{fdaoutlier} and \texttt{ddalpha}.
Since all outlier detection methods are able to benefit from transformations, e.g., computing derivatives of the original curve, (\citealp[cf.][Section~3.1]{NGH17}; \citet{DMSG20}), we consider only detection methods based on the original curves for simplicity.
{The third study compares the MSEs of medians based on the different functional depths mentioned above. All these methods are implemented in R packages \texttt{fdaoutlier} and \texttt{ddalpha}.}


\subsection{Setting up different types of outliers} \label{ssec_4_1}

To evaluate the performance of our proposed RHD, we consider inliers that are relatively smooth curves and different types of outliers, which include outliers in either magnitude or shape. The shape outliers are outlying only in terms of their pattern but not in magnitude. All curves are evaluated at 50 equally spaced time points in $[0,1]$.

Inlier curves $\xx_{n_{in}} = \{X_i\}_{i=1}^{n_{in}}$ 
are generated as
\begin{align} \label{eqKLtrunc}
	X = \sum_{j=1}^{J_0} \sqrt{\g_j} \xi_j \phi_j
\end{align}
with $J_0 = 15$,
where $\{\g_j\}_{j=1}^\infty$ is a non-increasing sequence of positive numbers with $\sum_{j=1}^\infty \g_j<\infty$ and $\{\phi_j\}_{j=1}^\infty$ is the set of trigonometric basis functions, which forms an orthonormal basis of $L^2([0,1])$. 
The random variables $\xi_j$ are chosen as $\xi_j \equiv \xi W_j$ where both $W_j$ and $\xi$ are independent $\mathsf{Unif}(-\sqrt{3}, \sqrt{3})$ random variables, so the random function $X$ is non-Gaussian with a bounded range.
The eigenvalues are set as $\g_j = 2\sum_{l=j}^\infty l^{-5}$, so that the eigengaps $\g_j - \g_{j+1} = 2j^{-5}$ follow a polynomial decay and give rise to relatively smooth inlier curves. 

\begin{figure}[b!]
	\centering
	\includegraphics[width=0.69\linewidth]{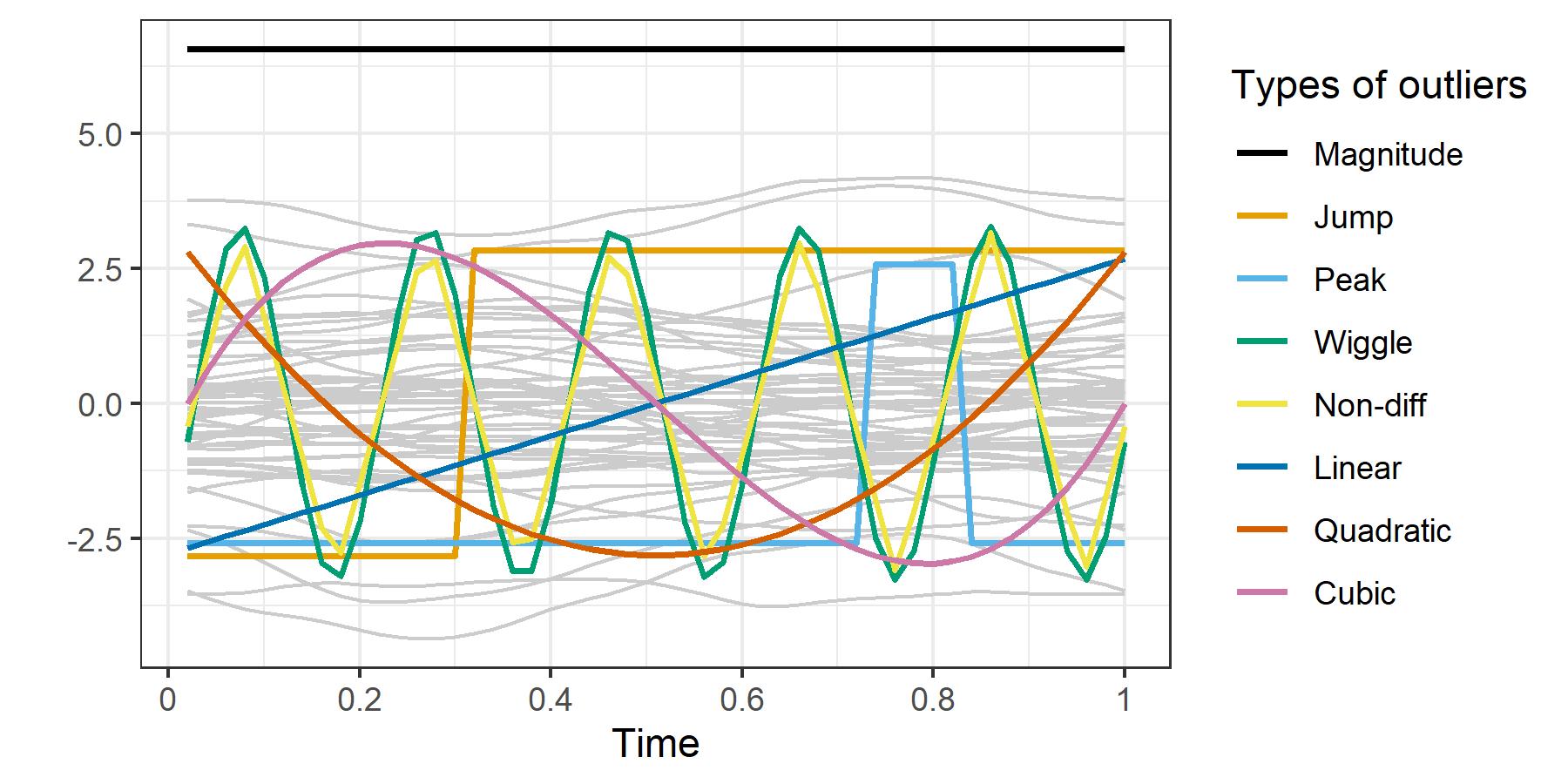}
	\caption{Illustration of the eight types of outliers (black and colored curves) and inliers (gray) considered in the simulation}
	\label{figSimDesign}
\end{figure}

Eight different types of outlier curves are considered, including one magnitude and seven shape outliers;	
\autoref{figSimDesign} displays a realization of the eight types of outliers
while Section~S2.2 in the supplement describes the details of their construction.
Note that the shape outliers are set to lie within the range of inliers at any time point, so outlier detection based on single time points will not perform well. 
Similar outliers have been described in \cite{DMSG20} and the references therein.


\subsection{Ranking outliers} \label{ssec_4_2}

We compute the RHD-based normalized rankings of a sample of curves as defined in \autoref{ssec_2_5}.
Each simulation scenario considers a sample contaminated by a different type of outlier.
We consider different sample sizes $n \in \{50, 200, 400\}$, truncation levels $J \in \{2, 4, 6, 8, 10\}$, 
and regularization parameters corresponding to quantile levels $u \in \{0.5, \dots, 0.9, 0.91, \dots, 0.99\}$.
We only present the results for $n=400$, $J=6$, and $u \in \{0.5, 0.7, 0.9, 0.95\}$,
and the supplement includes extra results and algorithmic details.

\begin{figure}[b!]
	\centering
	\includegraphics[width=0.7\linewidth]{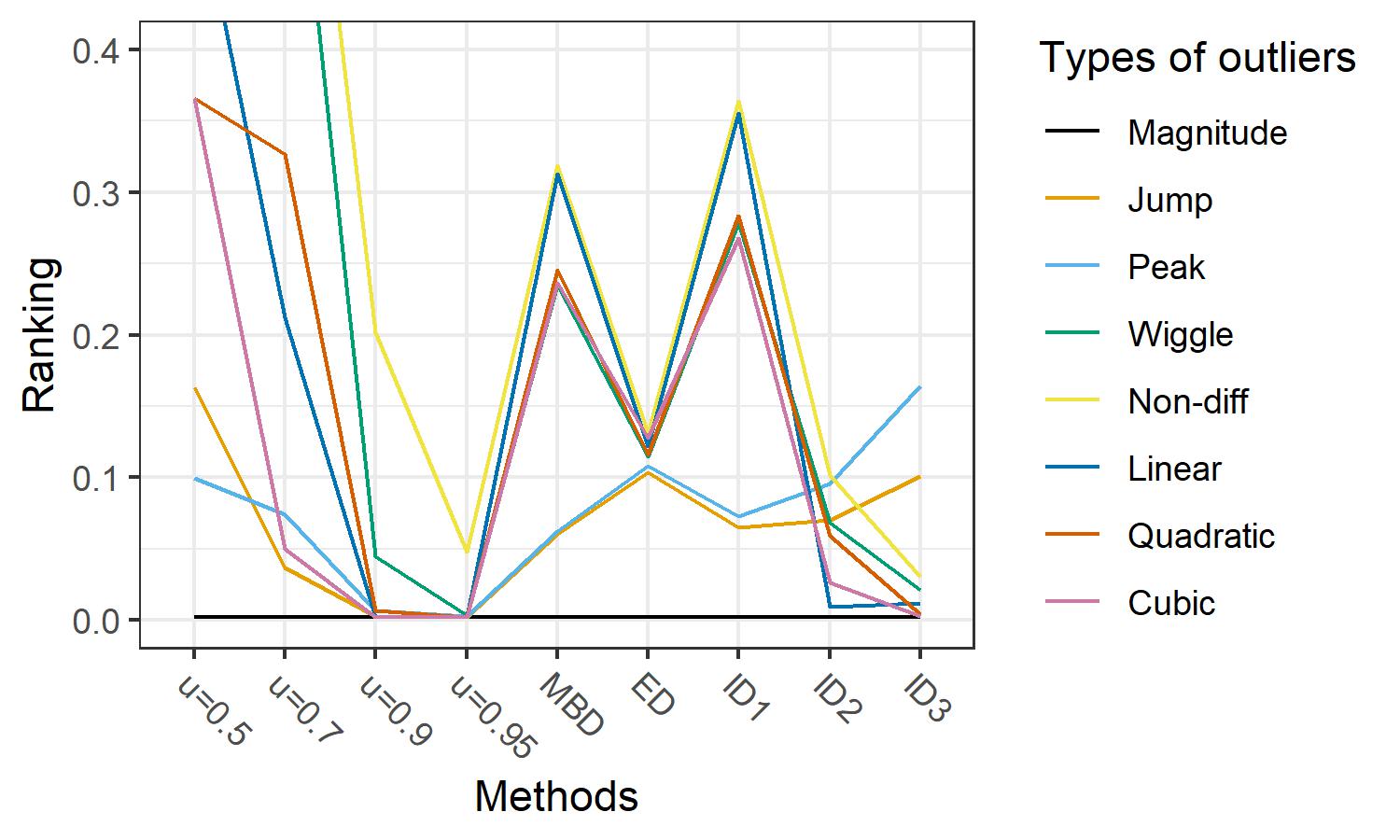}
	\caption{
		RHD-based normalized rankings of the eight outliers considered from different depth methods when $n=400$.  
		Lower rank is better since it shows that the outlier can be detected.
		For the proposed RHD, $J=6$ and  $u \in \{0.5, 0.7, 0.9, 0.95\}$  are considered.
	}
	\label{figSim2Rankingbyn}
\end{figure}

To evaluate the performance of RHD for ranking observations and flagging possible outliers, the normalized rankings of the outlier using different depth notions are compared in \autoref{figSim2Rankingbyn}.
The magnitude outlier always has the lowest depth using any depth notion, including the proposed RHD with the smallest regularization parameter ($u=0.5$)
which results in a depth sensitive to only magnitude but not shape outliers. 
In contrast, the shape outliers, which are harder to detect, have different rankings depending on the depth notions and the tuning parameter $\lambda$ for RHD. 
The proposed RHD assigns the smallest depth to most of the shape outliers when $\lambda$ is large, i.e., $u=0.9$ or $u=0.95$, 
except for the wiggle outlier when $u=0.9$ 
and the non-differentiable outlier when either $u=0.9$ or $u=0.95$. \
The MBD, ED, and the first order integrated depth, $\mathrm{ID}_1$, do not give the lowest rank to any of the shape outliers, while the second and third  order integrated depths, $\mathrm{ID}_2$ and $\mathrm{ID}_3$, 
perform quite well in general except for the scenarios with either jump or peak outliers, which have sharp features only in a small time window.

\subsection{Outlier detection} \label{ssec_4_3}

We next examine the performance of the proposed outlier detection method based on the RHD. 
The evaluation metrics are chosen to be the proportion $p_c$ of correctly detected outliers and the proportion $p_f$ of falsely detected outliers \citep{SG11, AR14, DMSG20} averaged over 1000 Monte Carlo experiments.
Again, only one type of outlier is added in the sample in each simulation scenario, and for brevity, we display only results for the same case $n=400$ as presented in \autoref{ssec_4_2}; 
see the Supplementary Materials for the complete results and details.
We calculate $p_c$ and $p_f$ for different factors $f \in \{1.5, 2.0, 2.5, 3.0, 3.5\}$ for functional boxplots and for the univariate boxplot in our proposed outlier detection method, 
but we do not vary the factor $f=1.5$ in the method by \cite{NGH17} since it is set fixed in the R package \texttt{ddalpha}. 

\begin{figure}[b!]
	\centering
	\includegraphics[width=0.99\linewidth]{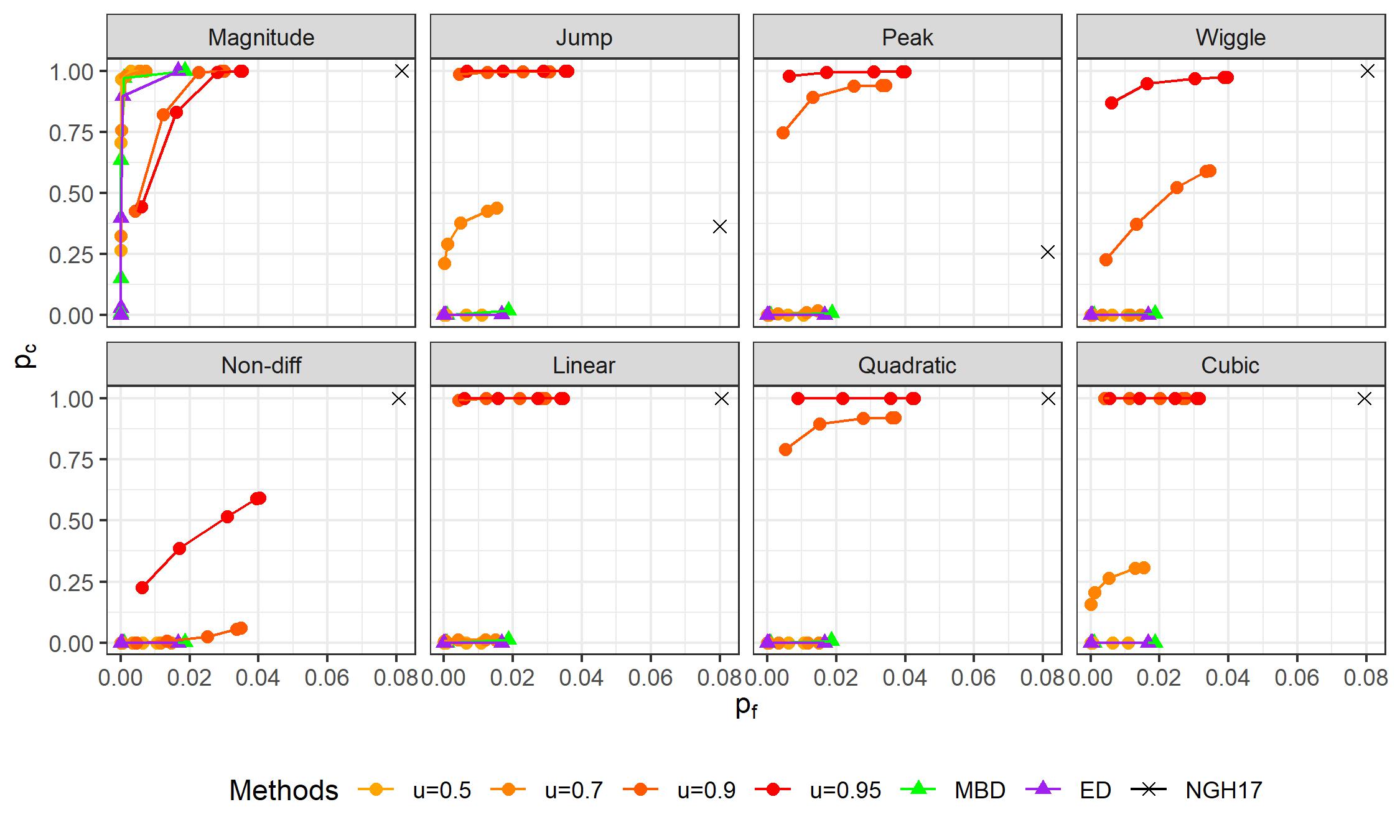}
	\caption{
		The ROC curves of the proportions of correctly versus falsely detected outliers based on 1000 Monte Carlo iterations. 
		Eight types of outliers are studied, one outlier is included in each dataset and the sample size is $n=400$.
	}
	\label{figSim3ROC1out}
\end{figure}	

\begin{figure}[b!]
	\centering
	\includegraphics[width=0.99\linewidth]{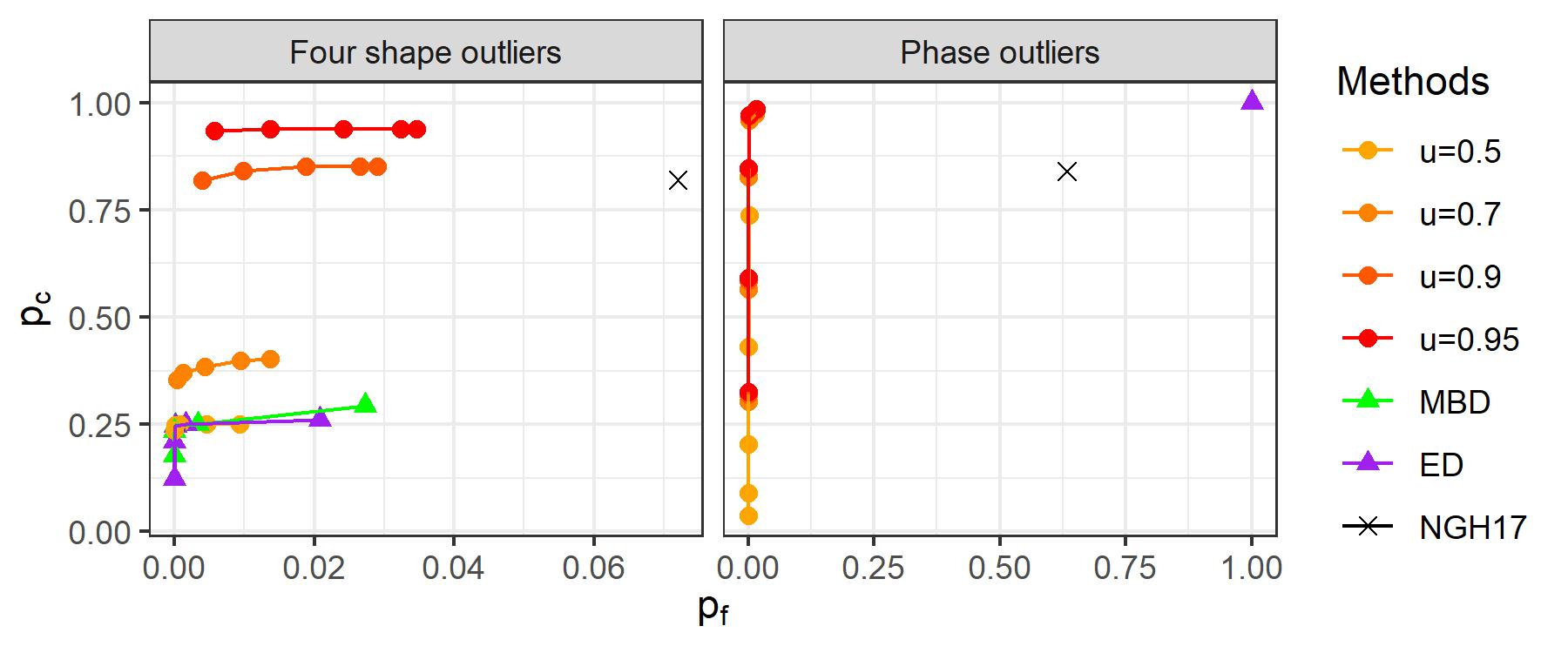}
	\caption{
		The ROC curves for different outlier detection methods.
		In the left panel, four shape outliers (magnitude, jump, wiggle, and linear outliers) are simultaneously included in each sample of size $n=200$.
		In the right panel, phase outliers described in Section~S2.5 are  included.
		Scales in x-axes are different. 
		}
	\label{figSim3ROC_mixed_wrap}
\end{figure}

{
In Figures~\ref{figSim3ROC1out}-\ref{figSim3ROC_mixed_wrap}, we present the results via the receiver operating characteristic (ROC) curve by drawing a scatterplot of $p_c$ versus $p_f$, where different points on a curve correspond to different factors $f$, and the color and shape of the dots in the lines indicate the outlier detection method used. 
The truncation level is set to be $J=6$ and four quantile levels $u \in \{0.5, 0.7, 0.9, 0.95\}$  
are considered for the proposed RHD,
and MBD, ED, and NGH17 respectively stand for functional boxplots based on either MBD or ED and the method by \cite{NGH17}, where the latter uses the integrated depths of the first three orders.
In these figures, closer to the upper-left corner indicates better performance.
}

{\autoref{figSim3ROC1out} shows the results when we only include one outlier of the types described in \autoref{figSimDesign}.}
First, for the magnitude outlier, our proposed method with smaller regularization parameter ($u=0.5$) performs very well and is comparable with the other existing methods. 
Second, for the shape outliers, the functional boxplots with either MBD or ED and our outlier detection method with $u=0.5$ (the smallest regularization parameter) are powerless in detecting the outliers. 
In contrast, under larger regularization $u=0.9$ or $u=0.95$, our proposed method mostly outperforms all the others.
If a larger quantile level such as $u=0.98$ is chosen, 
the correct detection rates $p_c$ of RHD-based method for non-differentiable outlier become higher and our method is competitive to the method by \cite{NGH17} in this setting; 
these additional results are given in the supplement.
The method proposed by \cite{NGH17} performs well in terms of correct detection rates $p_c$ except for the jump and peak outliers, 
but the false detection rates $p_f$ are relatively high. 
In summary, our proposed outlier detection method with larger regularization parameters exhibits high $p_c$ and low $p_f$ for all types of shape outliers.


We also show a more practical scenario where different types of outliers are simultaneously included in the sample. 
Four types of outliers are selected for this numerical study: magnitude, jump, wiggle, and linear outliers.
The ROC curve is displayed in the left panel of \autoref{figSim3ROC_mixed_wrap} for sample size $n=200$. 
The truncation level of the proposed RHD is again $J=6$ and the quantile levels for the regularization parameter $\ld$ are $u \in \{0.5, 0.7, 0.9, 0.95\}$.
In this case, the proposed outlier detection method with large regularization $u=0.9$ and $u=0.95$ are superior to the other methods in terms of both $p_c$ and $p_f$.

{
Inspired by the associate editor's feedback, we implemented an additional simulation study to evaluate the robustness of our method against phase outliers, which have equal magnitude and shape but different phase variation (cf.~\cite{AR12, MRSS15, XKBS17}). 
We adopt the third simulation set-up from \cite{XKBS17}, with the sample size $n=100$ and 10\% outliers;
the detailed data generation process is described in Section~S2.5. 
The right panel of \autoref{figSim3ROC_mixed_wrap} shows the ROC curve for this phase outlier scenario. 
The results there reveal that our method is superior in detecting phase outliers compared to the other methods. 
While the functional boxplots with MBD and ED fail to identify phase outliers since all the observations were considered outliers and the method by \cite{NGH17} still had fairly high $p_f$, our outlier detection method exhibits high $p_c$ and low $p_f$, particularly when a small adjustment factor $f$ is chosen and $u$ is large.
These findings underscore/emphasize the robustness and good performance of our proposed method in the presence of phase outliers relative to the other methods.
}


{

\subsection{Median estimation} \label{ssec_4_4}

We explore the performance of our method from a different perspective: depth-based medians.
We define the median element for the depth functions under consideration as the curve that attains the maximal depth value.
In cases of ties, we use the average of the deepest curves to determine the median.
In the simulation set-up outlined in \autoref{ssec_4_1}, we added a 10\% contaminated sample, with curves generated either as the magnitude and shape outliers depicted in \autoref{figSimDesign} or as the phase outliers described in Section~S2.5 of the supplement.
For each Monte Carlo iteration, we calculate the depth-based median and its $L^2$ distance from the center.
The average of all iterations is used to to approximate the MSEs of the resulting medians.
The detailed description of the simulation procedure can be found in Section~S2.6 of the supplement.
See \cite{DL23} for a similar simulation study. 
Tables~\ref{tb1}-\ref{tb2} present the MSEs of the depth-based medians in various contamination scenarios. 
The MSEs of the sample average are also reported for comparison.
The smallest MSEs among the RHD-based depths are emphasized in bold, while those among the other depths are italicized.

\autoref{tb1} provides the MSE results when the sample with size $n=200$ is contaminated by 10\% of magnitude or shape outliers. 
In these magnitude/shape contamination scenarios, since the inliers are generated following \eqref{eqKLtrunc}, which is halfspace symmetric about center $0 \in \HH$, we use zero element for the true center. 
The proposed RHD shows strong performance against most outliers, outperforming our main competitors MBD, ED, and ID$_1$, and showing comparable results with ID$_2$ and ID$_3$. This demonstrates the robustness of the RHD-median in the presence of various types of contamination.

\begin{table}[b!]
	\centering 
	\caption{
		The mean squared errors (MSEs) of the depth-based median from different depth functions. The sample size is $n=200$ with 10\% contaminated sample. 
		}
	\label{tb1}
	\small \setstretch{1.3}
	\begin{tabular}{c|cccccccccc}
		\hline
		Methods & $u=0.5$ & $u=0.7$ & $u=0.9$ & $u=0.95$ & MBD & ED & ID$_1$ & ID$_2$ & ID$_3$ & $\bar{X}$ \\ \hline
		Pure &  4.54 &  3.94 &  2.65 &  \textbf{2.63} &  
		3.68 &  3.80 &  3.71 &   2.02 &   \textit{1.43} &   9.84 \\ 
		Magnitude & 21.83 & \textbf{20.55} & 21.79 & 21.54 & 
		19.30 & 19.37 & 19.44 &  11.01 &   \textit{5.42} & 345.88 \\ 
		Jump &  5.08 &  4.83 &  4.35 &  \textbf{3.95} &  
		7.93 &  6.70 &  8.33 &   2.37 &   \textit{1.74} &  50.50 \\ 
		Peak & 20.34 & 19.29 &  7.02 &  \textbf{4.96} & 
		13.85 & 13.72 & 13.76 &   8.02 &   \textit{4.31} &  65.20 \\ 
		Wiggle &  5.27 &  3.71 &  2.96 &  \textbf{2.82} &  
		3.98 &  3.99 &  3.97 &   2.33 &   \textit{1.83} &  26.67 \\ 
		Non-diff &  5.41 &  4.00 &  2.94 &  \textbf{2.74} &  
		3.87 &  4.18 &  3.78 &   2.40 &   \textit{1.80} &  31.33 \\ 
		Linear &  5.55 &  5.41 &  5.05 &  \textbf{4.55} & 
		13.15 &  6.42 & 14.96 &   3.23 &   \textit{1.85} &  37.53 \\ 
		Quadratic &  5.36 &  4.41 &  3.75 &  \textbf{3.62} &  
		9.46 &  4.89 & 13.22 &   3.13 &   \textit{2.24} &  48.60 \\ 
		Cubic &  5.46 &  5.43 &  5.20 &  \textbf{5.16} & 
		13.56 &  6.84 & 15.48 &   4.05 &   \textit{2.29} &  50.52 \\  \hline
	\end{tabular}
	
\end{table}

We conducted an extra median simulation study including contaminated samples with phase variation.
We again adopted the third simulation set-up from \cite{XKBS17}, and introduced an additional scenario with stronger phase variation.
This additional case (denoted as ``Strong'') features a stronger phase variation in the outliers than the existing set-up in \cite{XKBS17} (denoted here as ``Weak''); see Figure~S6 in the supplement for an illustration of curves generated from these strong and weak variations settings.
This allows us to compare the robustness of the depth-based medians against changes in the strength of the phase contamination.

\begin{table}[b!]
	\centering 
	\caption{
		The mean squared errors (MSEs) of the depth-based median from different depth functions. The sample size is $n=100$ with 10\% contaminated sample in phase variation; see Section~S2.5 of the supplement for the data generation. 
		}
	\label{tb2}
	\small \setstretch{1.3}
	\begin{tabular}{c|cccccccccc}
		\hline
		Methods & $u=0.5$ & $u=0.7$ & $u=0.9$ & $u=0.95$ & MBD & ED & ID$_1$ & ID$_2$ & ID$_3$ & $\bar{X}$ \\ \hline
		Weak & \textbf{18.63} & 31.96 & 32.06 & 31.96 & 
		\textit{17.58} & 41.59 & 17.59 &  18.11 &  46.49 &  23.57 \\ 
		Strong & 72.83 & 45.63 & 38.89 & \textbf{38.68} 
		& \textit{16.57} & 39.09 & 17.49 & 109.06 & 163.81 &  31.1 \\ \hline
	\end{tabular}
	
\end{table}

\autoref{tb2} displays the MSEs of the depth-based medians under phase contamination;
the detailed data generation procedure and the true center location are given in the supplement. 
For weak phase variation, the MBD, ID$_1$, and the RHD with $u=0.5$ perform best
while the RHD with higher regularization parameters show moderate performance. 
With strong phase variation, the MBD and ID$_1$ continue to outperform the others,
while the RHD with higher regularization parameters maintain moderate performance. 
Interestingly, the MSEs of medians from ID$_2$ and ID$_3$ are extremely large with strong phase contamination,
suggesting that ID$_2$ and ID$_3$ are not robust against phase contamination. 
In contrast, the RHD-based median with high quantile levels yield similar MSE with moderate size regardless of the strength of phase variation, 
although they do not exhibit the best performance for either phase variation.
Overall, we conclude that each depth-based median has their own relative merits and there is no depth that outperforms all the others in every scenario. However, the RHD-based median performs well in general with less sensitivity to the types of outliers.

}




\section{Data Applications} \label{sec5}

We apply the RHD-based outlier detection method to real data that exhibit very different characteristics:
medfly, world population growth (WPG), and sea surface temperature (SST) datasets.
The curves in the medfly example are very dynamic and complex, while the other two datasets consist of smoother and relatively simpler curves that are well-represented by the first few eigenfunctions.
For each data example, four regularization parameters $\ld_1 < \ld_2 < \ld_3 < \ld_4$ based on the quantile levels $u \in  \{0.4, 0.6, 0.8, 0.95\}$ and five adjustment factors $\{1.5, 2.0, 2.5, 3.0, 3.5\}$ are considered. 
In all illustrated figures, outliers are marked in red, while inliers are denoted as dashed grey curves. The deepest curve is marked in blue for reference; if there are multiple deepest curves, we display their average.
In this section, we report only the results from the analysis of the medfly dataset while the other results regarding the WPG and SST datasets are deferred to the supplement.

\begin{figure}[b!]
	\centering
	\includegraphics[width=0.79\linewidth]{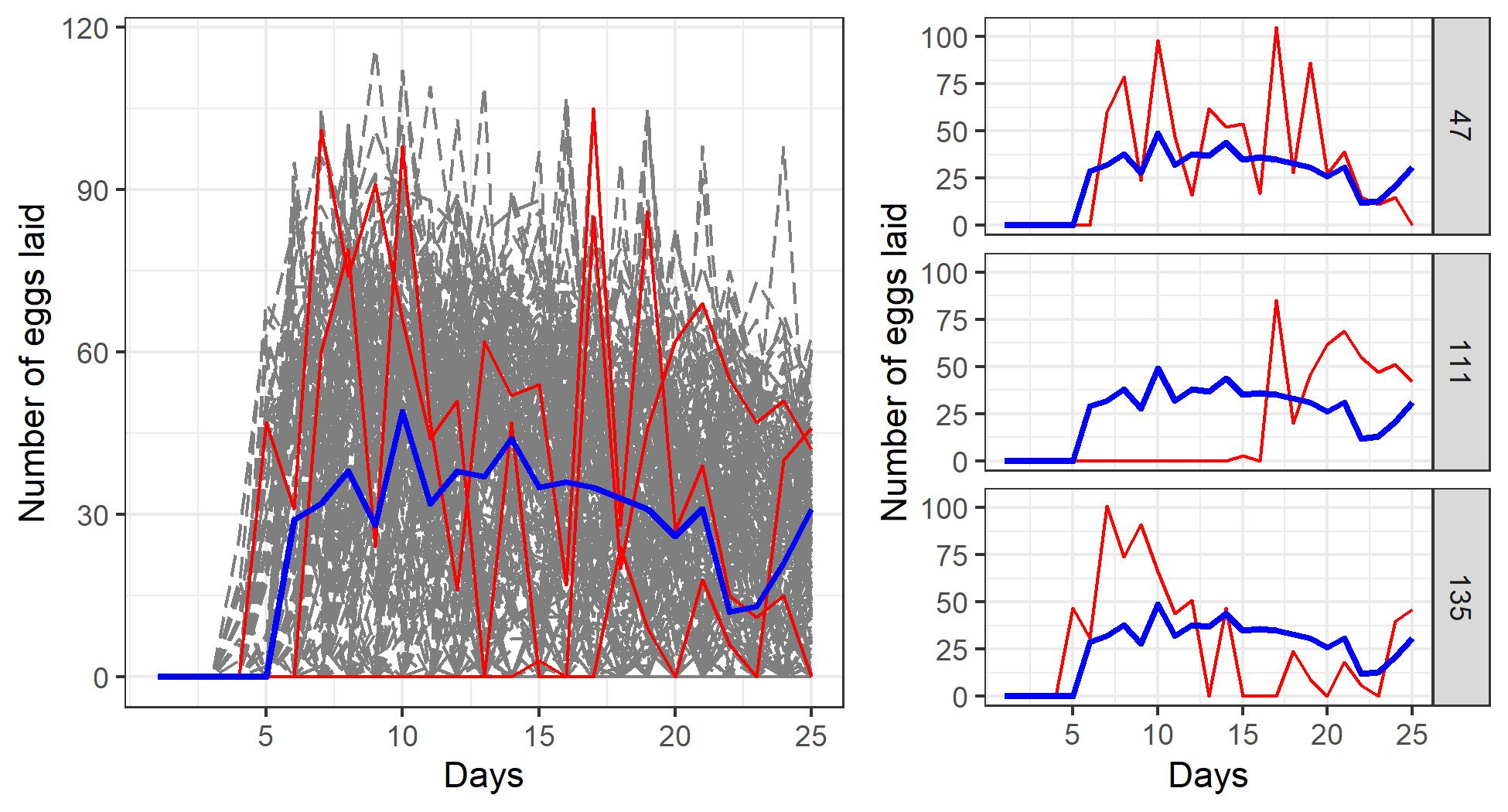}
	\caption{
		The medfly dataset, showing the outlying curves (red) with IDs 47, 111, and 135 detected by the proposed RHD outlier detection method with the truncation $J=8$, adjustment factor $f=3.5$, and regularization $\ld = \ld_4 = 0.0507$ (from the quantile level $u=0.95$). 
		All 200 data curves are exhibited in the left panel with the outliers represented in red and the deepest trajectory in blue.
		The three flagged outliers are separately displayed in the three sub-panels on the right. 
	}
	\label{figRDA3_medfly_combined}
\end{figure}

The medfly dataset contains daily observations of number of eggs laid by female Mediterranean fruit flies (medflies, \textit{Ceratitis capitata}) described in \cite{carey98}.
We analyze a subset of $n=200$ egg-laying trajectories in the first 25 days after hatch (see \autoref{figRDA3_medfly_combined}), randomly selected from a preprocessed dataset containing 789 medflies available in the R package \texttt{fdapace}.

Due to the complexity of the medfly trajectories, the functional boxplots constructed with either MBD  \citep{LR09} or ED \citep{NN16} detect all 200 curves as outliers. Note that MBD and ED are originally defined for smooth functions and might not be appropriate for analyzing very irregular and non-smooth curves. 
Functional boxplot flags a curve as outlier if it lies outside of the whisker band at any time point, but this does not make sense for this wiggly data because medflies tend to attain very high or low productivity at some random time in their life.
In addition, the functional boxplot is hard to read when the curves are often oscillating. 

Hence, for this type of data, alternative outlier detection methods are needed to reflect the high variability in the sample. The proposed RHD is defined in the space of square integrable functions, a larger class than the space of smooth functions.
Furthermore, our outlier detection method is data-adaptive and measures the outlyingness of each curve depending on the extreme patterns inherent in the samples.	
Therefore, the proposed outlier detection method based on RHD is adequate to deal with such complex structured curves and in fact performs well as shown next.

\autoref{figRDA3_medfly_combined} displays the outlier detection results from the proposed method with the truncation level $J=8$ and the adjustment factor $f=3.5$;
{the first eight FPCs (chosen by truncation $J=8$) can explain 87.61\% of total variability in the observed curves}
while the adjustment factor $f=3.5$ is determined by Algorithm~S1 in the supplement. Only the results from $\ld = \ld_4 = 0.0507$ are reported here since the method detects no outliers when $\ld \in \{\ld_1, \ld_2, \ld_3\}$. 
The proposed method flags three outliers denoted by their IDs in the subsample, 47, 111, and 135. 
To see their shapes more clearly, we display one outlier in each panel (along with the deepest curve) in the right side of \autoref{figRDA3_medfly_combined}.
The trajectory for the outlying Medfly 47 is more wiggly than the inliers; Medfly 111 appears to be outlying due to its late start of otherwise high fertility; Medfly 135 has a relatively narrow peak of fertility from day 7 to 12 and then
stays lower than most inliers thereafter.
These results demonstrate that the RHD and the associated outlier detection method are able to handle non-smooth trajectories without presmoothing and reveal shape outliers with distinct patterns.


\section*{Supplementary material}
The supplement includes technical details, extra simulation studies, and additional results regarding data applications.

\section*{R package}

The R-package consists of the following two functions: a function that computes the RHD values and implements the proposed outlier detection procedure as described in Algorithms~\ref{alg1}-\ref{alg2}; 
a function that selects the adjustment factor as described in Algorithm~S1 in the supplement.

%

%
%
%
%
%
%

\bibliographystyle{dcu}
\bibliography{RHDpreprint_jrssb_bibfile}

\end{document}